\documentclass[final,authoryear,1p,times]{elsarticle}

 \usepackage{graphicx}

\usepackage{amssymb}

\usepackage{amsmath}
\usepackage{multirow}
  \IfFileExists{amsbsy.sty}
    {\typeout{^^JFound the 'amsbsy' package on the system, using it.^^J}%
     \usepackage{amsbsy}}
    {\providecommand\boldsymbol[1]{\mbox{\boldmath $##1$}}}
\providecommand\bnabla{\boldsymbol{\nabla}}
\newcommand{\Rm}{\mbox{Rm}}
\newcommand{\Rey}{\mbox{Re}}

\journal{Comptes Rendus Physique}

\begin{document}

\begin{frontmatter}



\title{Magnetic induction maps \\
in a magnetized spherical Couette flow experiment}


\author{Henri-Claude Nataf}
\ead{Henri-Claude.Nataf@ujf-grenoble.fr}
\address{ISTerre, Universit\'e de Grenoble 1, CNRS, F-38041 Grenoble, France}

\begin{abstract}
The $DTS$ experiment is a spherical Couette flow experiment with an imposed dipolar magnetic field.
Liquid sodium is used as a working fluid.
In a series of measurement campaigns, we have obtained data on the mean axisymmetric velocity, the mean induced magnetic field and electric potentials.
All these quantities are coupled through the induction equation.
In particular, a strong $\omega$-effect is produced by differential rotation within the fluid shell, inducing a significant azimuthal magnetic field.
Taking advantage of the simple spherical geometry of the experiment, I expand the azimuthal and meridional fields into Legendre polynomials and derive the expressions that relate all measurements to the radial functions of the velocity field for each harmonic degree.
For small magnetic Reynolds numbers $\Rm$ the relations are linear, and the azimuthal and meridional equations decouple.
Selecting a set of measurements for a given rotation frequency of the inner sphere ($\Rm \simeq 9.4$), I invert simultaneously the velocity and the magnetic data and thus reconstruct both the azimuthal and the meridional fields within the fluid shell.
The results demonstrate the good internal consistency of the measurements, and indicate that turbulent non-axisymmetric fluctuations do not contribute significantly to the axisymmetric magnetic induction.

\quad

\noindent \textbf{R\'esum\'e}

\quad

\textbf{Cartographie de l'induction magnŽtique dans un \'ecoulement de Couette sph\'erique soumis \`a un champ magn\'etique.}
L'exp\'erience $DTS$ consiste en un \'ecoulement de Couette sph\'erique soumis \`a un champ magn\'etique dipolaire.
Le fluide utilis\'e est du sodium liquide.
Au cours d'une s\'erie de campagnes de mesure, nous avons obtenu des donn\'ees sur le champ de vitesse moyen axisym\'etrique, le champ magn\'etique moyen, et le potentiel \'electrique.
Toutes ces quantit\'es sont coupl\'ees \`a travers l'\'equation d'induction.
En particulier, la rotation diff\'erentielle du fluide produit un fort effet $\omega$ qui induit un champ magn\'etique azimutal cons\'equent.
Profitant de la g\'eom\'etrie sph\'erique de l'exp\'erience, je d\'eveloppe les champs azimutaux et m\'eridionaux en polyn\^omes de Legendre et j'obtiens les expressions qui relient toutes les mesures aux fonctions radiales du champ de vitesse pour chaque degr\'e.
Pour de petits nombres de Reynolds magn\'etiques $\Rm$ les relations sont lin\'eaires et les \'equations azimutale et m\'eridionale sont d\'ecoupl\'ees.
Je s\'electionne un jeu de mesures pour une vitesse de rotation donn\'ee de la sph\`ere interne ($\Rm \simeq 9.4$) et j'inverse simultan\'ement les donn\'ees de vitesse et magn\'etiques, reconstruisant ainsi \`a la fois les champs azimutaux et m\'eridionaux dans la coquille fluide.
Les r\'esultats d\'emontrent la bonne coh\'erence des mesures et indiquent que les fluctuations turbulentes non-axisym\'etriques ne contribuent pas de fa\c con significative \`a l'induction magn\'etique axisym\'etrique.

\end{abstract}

\begin{keyword}
dynamo \sep magnetohydrodynamics \sep omega effect \sep liquid sodium \sep DTS

\quad

{\textit Mots-cl\'es : } magn\'etohydrodynamique \sep effet om\'ega \sep sodium liquide \sep DTS

\end{keyword}

\end{frontmatter}


\section{Introduction}
\label{Introduction}

It is now well established that the magnetic field of most planets and stars is generated by the dynamo mechanism \citep{larmor19, elsasser46a, parker55}, in which motions within an electrically conducting medium amplify infinitesimally small magnetic field fluctuations up to a level where the Lorentz force that results is large enough to stop their amplification.
This is possible for large enough values of the magnetic Reynolds number $\Rm = U L/\eta$ (where $U$ is a typical flow velocity, $L$ a typical length, and $\eta$ is the magnetic diffusivity of the medium).

After the success of the first fluid dynamo experiments in Riga \citep{gailitis01, gailitis08} and Karlsruhe \citep{stieglitz01, mueller08}, it was felt that the next step was to obtain dynamo action in a highly turbulent free-to-adjust flow.
Indeed, the flow was very much constrained by walls and pipes in those pioneering liquid sodium experiments, while the flow has much more freedom in natural dynamos, and the amplitude of turbulent fluctuations was less than 10\% that of the mean flow, while it can be of order one or more in astrophysical objects.

The challenge was soon addressed by several teams, in Russia \citep{frick01}, in the USA \citep{lathrop01, oconnell01}, and in Europe \citep{cardin02, marie02}, all using liquid sodium as a working fluid (see recent reviews by \citet{verhille10} and \citet{lathrop11}).
In several of these experiments, the mean flow was well characterized, and was expected to yield a dynamo above a critical magnetic Reynolds number that was achievable.
It therefore came as a surprise that none of these experiments produced a dynamo, although a rich variety of dynamo behaviours have been discovered in the $VKS$ experiment in Cadarache, France \citep{berhanu07, monchaux07, aumaitre08} when ferromagnetic disks stir the fluid.

It appears that turbulent fluctuations have a collective contribution to the mean magnetic induction that counteracts that of the mean flow \citep{spence06}.
That small-scale turbulent fluctuations can contribute to a large-scale magnetic field is not surprising.
In fact, it is the basis of the $\alpha$-effect introduced by \citet{steenbeck66}, who showed that the interaction of small-scale velocity fluctuations with small-scale magnetic field fluctuations produce an electromotive force proportional to a large-scale magnetic field.
The success of the Karlsruhe dynamo experiment relies precisely on this collective effect.

In rotating bodies such as planets and stars, it has long been recognized that the combination of an $\alpha$-effect with an $\omega$-effect provides an appropriate recipe to produce a large-scale magnetic field \citep{parker55} (see \citet{charbonneau05} and \citet{rieutord08} for reviews).
The $\omega$-effect is due to the shear caused by differential rotation in the fluid.
It converts poloidal magnetic field into toroidal magnetic field.
One then invokes some kind of $\alpha$-effect, due to non-axisymmetric motions, to convert toroidal field into poloidal field, in order to sustain the dynamo.
Many mean-field dynamo models for the Earth or the Sun are based on this mechanism \citep[e.g.,][]{parker55, barenghi91, brandenburg91, charbonneau05}.

It is therefore of some interest to evaluate these two effects in various situations, in particular to understand when, where and why turbulent fluctuations favor or hinder dynamo action.
There has been several efforts in this direction, using either numerical simulations \citep{spence08, brandenburg10}, or laboratory experiments \citep{petrelis03, spence06}, by observing the magnetic response of the device when a weak external magnetic test field is applied.
The $VKS$ team in particular has obtained clear evidence for these effects (and others) in von K\'arm\'an flows, as nicely summarized by \citet{verhille10}.
In a similar geometry, the Madison group has measured a dipolar component of the induced magnetic field, which cannot be explained by the mean flow \citep{spence06}.
More recently, a direct measurement of the $\alpha$-effect has been obtained on the same set-up \citep{rahbarnia12}, using a dedicated probe that measures locally the electromotive force.
Turbulent fluctuations can also enhance the magnetic diffusivity of the fluid, yielding the so-called $\beta$-effect.
Impressive measurements of this effect have been obtained by the group in Perm \citep{frick10, noskov12} for the transient flow of sodium in a torus abruptly stopped.
In these studies, only a global averaged response of the system is obtained, with no information on the spatial distribution of the $\alpha$- and $\omega$-effects.

Several teams have also been able to combine experimental measurements of the induced field with numerical simulations of the flow backed by velocity measurements on a water model of the experiment \citep{spence08, ravelet12}.
The present article presents one effort in the same direction, using the $DTS$ magnetized spherical Couette flow experiment.
The $DTS$ experiment is the only experiment so far that combines rotation and a strong applied dipolar magnetic field.
It was built to explore the resulting magnetostrophic regime \citep{cardin02}.
A number of interesting observations have already been presented, relating to the mean flow \citep{nataf06, nataf08a, brito11} and to the fluctuations \citep{schmitt08, schmitt12}.
Since the imposed magnetic field governs the dynamics of the system, it is not possible to rely on a water model.
However, we already know that turbulent fluctuations are strongly affected by the presence of a large scale magnetic field \citep{nataf08, schmitt12, figueroa12}, and it would be most interesting to find out how this translates into the $\alpha$-, $\beta$- and $\omega$-effects for this kind of flow.
Fortunately, \citet{brito11} have shown that velocity profiles measured by {\it in situ} ultrasound Doppler velocimetry provide excellent constraints of the mean flow in $DTS$.
Furthermore, the simple spherical geometry of this experiment makes it possible to use tools developed in the context of celestial dynamos.

Thus, my aim here is to set the stage for further steps that should enable us to map the $\alpha$-, $\beta$- and $\omega$-effects for the magnetized spherical Couette flow, using measurements from the $DTS$ experiment.
The first step consists in solving the kinematic problem of the large-scale fields, ignoring the contribution of the small-scale fluctuations.
For future reference, I write down the equations and the inversion procedure that apply when only the large-scale fields are taken into account.
I then apply this procedure to a selected set of actual $DTS$ measurements.
The results demonstrate the good internal consistency of the measurements and show that small-scale fluctuations play a negligible role.

The organization of this article is as follows: Section \ref{Experimental} describes the $DTS$ experiment, the measurements and the way they are processed.
Section \ref{Field decomposition} explains why we treat separately the azimuthal part and the meridional part of the fields.
The forward problem of predicting the induced field and related quantities from the mean flow is treated in section \ref{Azimuthal forward problem} for the azimuthal fields, and in section \ref{Meridional forward problem} for the meridional fields.
Section \ref{Azimuthal inversion} presents the inversion results for the azimuthal fields, and section \ref{Meridional inversion} for the meridional fields.
Conclusions and perspectives are discussed in section \ref{Discussion}.

\section{Experimental}
\label{Experimental}

\subsection{Experimental set-up}
\label{Experimental set-up}

\begin{figure}
	\begin{center}
		\includegraphics[width=8cm]{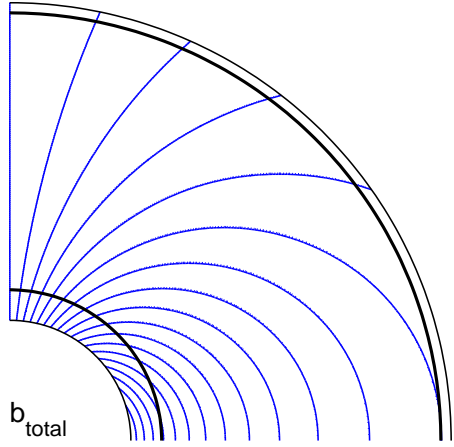}
	\caption{Sketch of the $DTS$ set-up.
In this northern quadrant of a meridional cross-section, liquid sodium is enclosed between the inner and outer solid shells.
The inner shell is made of copper and extends from radius $\hat{r}_i$ (thin line) to $r_i$ (thick line).
It contains a strong magnet, which produces an axial dipolar magnetic field, as shown by the blue field lines.
The fluid is entrained by the rotation of the inner sphere at frequency $f$ around the vertical axis.
The outer shell is made of stainless steel and extends from radius $r_o$ (thick line) to $\hat{r}_o$ (thin line).
It is at rest in the laboratory frame.
}
	\end{center}
	\label{fig:set-up}
\end{figure}

The $DTS$ experiment is a spherical Couette flow experiment with an imposed dipolar magnetic field.
Liquid sodium is used as a working fluid.
It is sketched in figure \ref{fig:set-up}.
It is contained between an inner sphere and a concentric outer shell, from radius $r_i^* = 74$ mm to $r_o^* = 210$ mm (the superscript $^*$ refers to the dimensional quantities, and is dropped after adimensionalization by $r_o^*$ as given in \ref{adimensionalization}).
The inner sphere consists of a $15$ mm-thick copper shell (dimensionless radius $\hat{r}_i$ to $r_i$), which encloses a permanent magnet that produces the imposed dipolar magnetic field $\boldsymbol{B_d}$, whose intensity reaches 175 mT at the equator of the inner sphere.
The stainless steel outer shell (dimensionless radius  $r_o$ to $\hat{r}_o$) is $5$ mm thick.
The inner sphere can rotate around the vertical axis (which is the axis of the dipole) at rotation rates $f=2 \pi \Omega$ up to $30$ Hz.
Although the outer shell can also rotate independently around the vertical axis in $DTS$, I only consider here the case when the outer sphere is at rest.
Additional details can be found in \citet{nataf08a} and \citet{brito11}.

\subsection{Dimensionless numbers}
\label{Dimensionless numbers}

In this article, I concentrate on the magnetic induction equation alone, which is governed by a single dimensionless number: the magnetic Reynolds number $\Rm$, which measures the induction over diffusion ratio.
However, the actual flow depends of course on other dimensionless numbers that determine the relative importance of forces in the Navier-Stokes equation.
Three independent dimensionless numbers can be constructed, and we choose the magnetic Reynolds number $\Rm$, the Elsasser number $\Lambda$ and the magnetic Prandtl number Pm, whose expressions are given in table \ref{tab:numbers}, together with their values in the $DTS$ experiment for rotation rates $f=3$ Hz and $f=30$ Hz.

\begin{table}[h] 
 	\begin{center}
		\begin{tabular}{ccccccc}
			$f$ & $\Rm$ & $\Lambda$ & Pm & \multicolumn{1}{|c}{$\Rey$} & Ha & Lu \\
			Hz & ${\Omega(r_o^*)^2}/{\eta}$ & ${\sigma B_0^2}/{(\rho \Omega)}$ & ${\nu}/{\eta}$ & \multicolumn{1}{|c}{${\Omega(r_o^*)^2}/{\nu}$} & ${r_o^* B_0}/{\sqrt{\rho \mu_0 \nu \eta}}$ & ${r_o^* B_0}/{\sqrt{\eta^2 \rho \mu_0}}$ \\
			\hline
			$3$ & $9.4$ & $3 \, 10^{-2}$ & \multirow{2}{*}{$7.4 \, 10^{-6}$} & \multicolumn{1}{|c}{$1.3 \, 10^{6}$} &  \multirow{2}{*}{$200$} &  \multirow{2}{*}{$0.5$} \\
			$30$ & $94$ & $3 \, 10^{-3}$ & & \multicolumn{1}{|c}{$1.3 \, 10^{7}$} &  & \\
		\end{tabular}
		\caption{\label{tab:numbers} Relevant dimensionless numbers and their values for two rotation rates $f$ of the inner sphere.}
	\end{center}
\end{table}

The magnetic Prandtl number simply compares the kinetic and magnetic diffusivities, and is a physical property of the fluid used (here liquid sodium).
The Elsasser number has been introduced to compare the Lorentz and the Coriolis forces.
It looks small in table \ref{tab:numbers} because we use $B_0$, the intensity of the imposed magnetic field at the equator of the outer sphere, to scale the magnetic field, but one should keep in mind that the actual imposed magnetic field is 23 times larger at the equator of the inner sphere.

The three other dimensionless numbers listed in table \ref{tab:numbers} can be deduced from the previous ones.
The Hartmann number Ha indicates that, even at the outer boundary, magnetic forces dominate over viscous forces.
The Lundquist number Lu compares the time-scale of Alfv\'en waves to their life-time.
It is needed to convert magnetic energies to the same units as kinetic energies.

\subsection{Data processing}
\label{Data processing}
We want to investigate the interaction of the flow with the magnetic field.
We thus want to combine measurements of the velocity field and of the induced magnetic field.
\citet{brito11} have recently shown that it was possible to retrieve the large scales of the mean axisymmetric velocity field in $DTS$ by inverting criss-crossing ultrasound Doppler velocity profiles.
The induced magnetic field is measured at several latitudes on the outer sphere.
Since the toroidal magnetic field vanishes at the surface, it is crucial to also measure the induced magnetic field inside the fluid.
Additional information on the interaction of the flow with the magnetic field is obtained by measuring the electric potential at the surface of the outer sphere, and the torque applied by the fluid on the spinning inner sphere.
\citet{brito11} present these various measurements and display their evolution with the rotation frequency $f$ of the inner sphere up to $30$ Hz.
Here I concentrate on the case with $f = \pm 3$ Hz ($\Rm \simeq 9.4$), for which we have the best velocity data coverage.
I give below further indications on the processing of the data I invert.

\subsubsection{ultrasound Doppler velocity profiles}
\label{Doppler}

\begin{figure}
	\begin{center}
		\includegraphics[width=6cm]{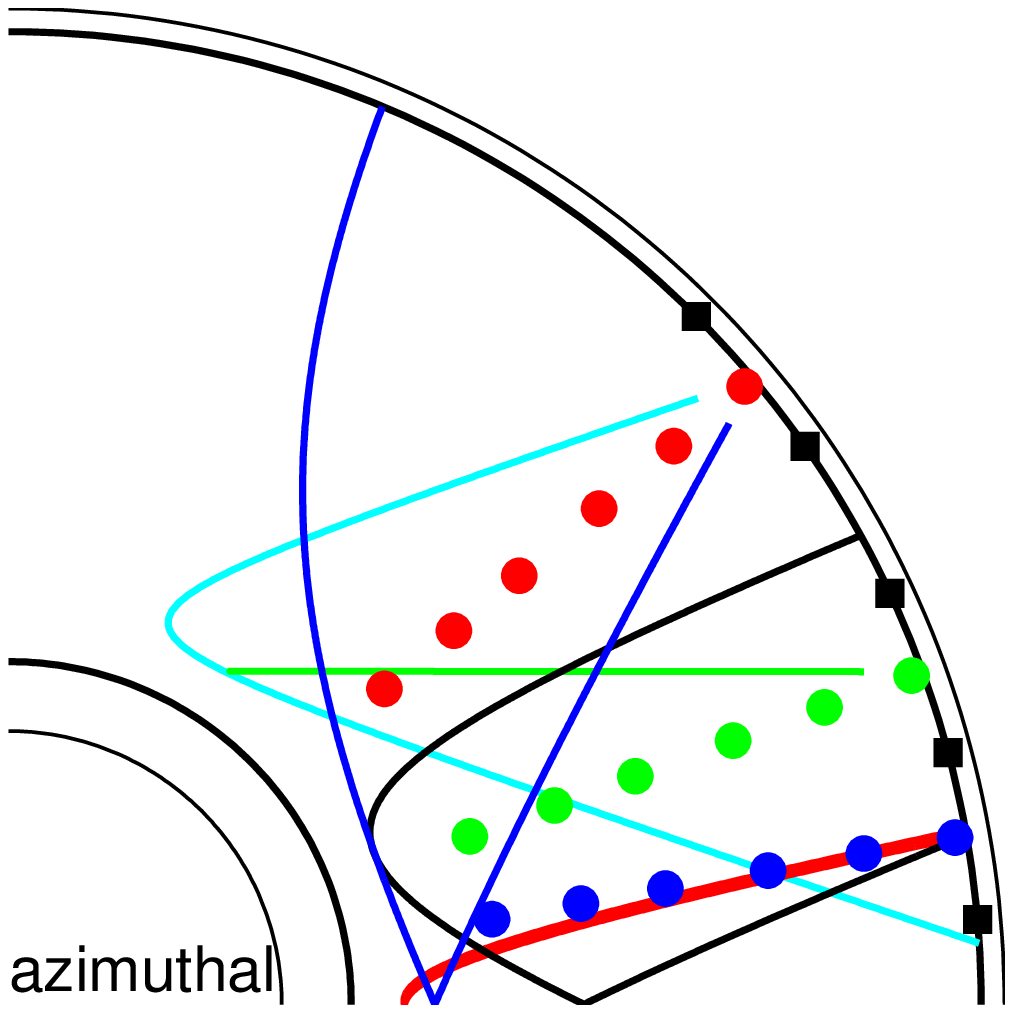}
		\includegraphics[width=6cm]{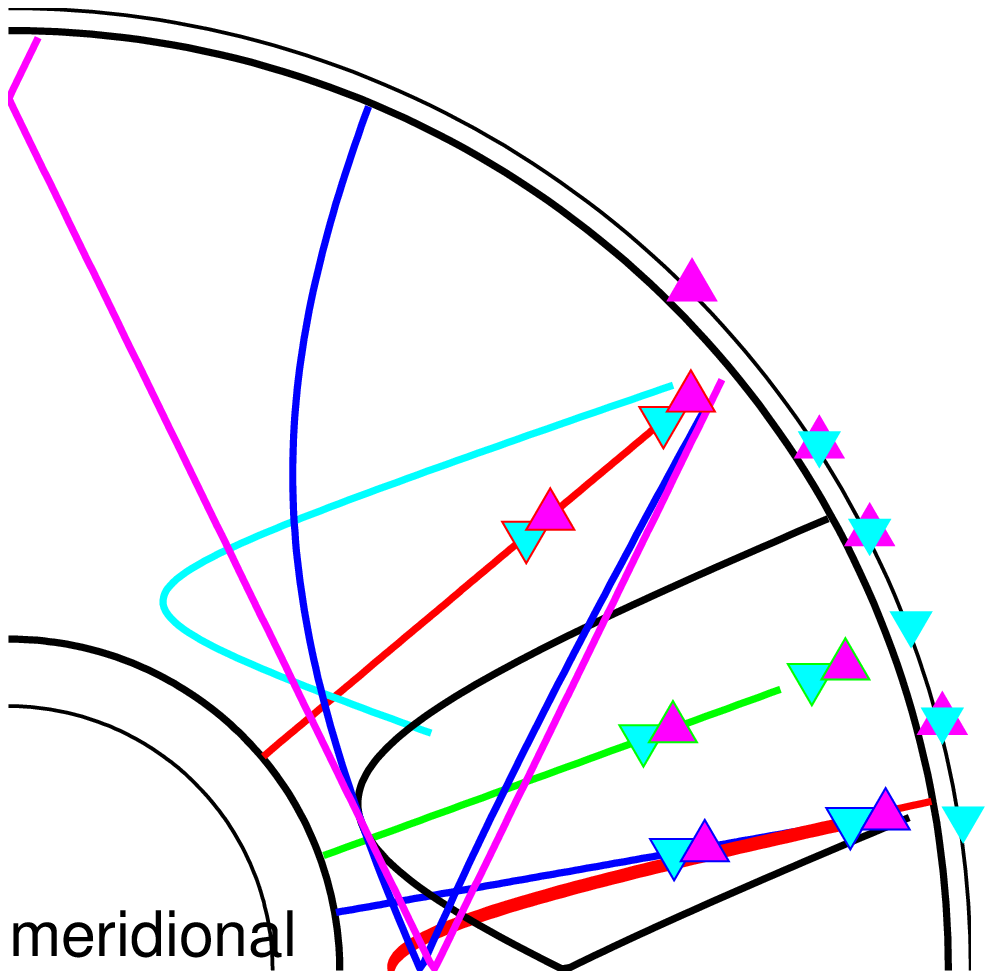}
	\caption{Data coverage in the northern quadrant of the ($s$,$z$) plane ($s$: cylindrical radius, $z$: vertical axis).
{\it left:} azimuthal fields. 
The colored curves are the projections in the ($s$,$z$) plane of the off-meridian ultrasound beams used to measure the azimuthal velocity of the fluid.
The colored dots are the positions where the azimuthal magnetic field is measured.
The electric potential is measured in the outer shell at latitudes given by the black squares.
{\it right:} meridional fields. The straight lines are radial ultrasound shots. The broken straight line is a meridional shot, and the curved lines are the projections in the ($s$,$z$) plane of the off-meridian shots.
The radial magnetic field is measured at the cyan triangles both inside the fluid and at the surface.
The orthoradial magnetic field data are from the magenta triangles.
}
	\end{center}
	\label{fig:data}
\end{figure}

Ultrasound Doppler velocimetry provides measurements of the component of the fluid velocity along the shooting line of the ultrasound beam, from the Doppler shift of back-scattered acoustic energy by small heterogeneities in the fluid.
In the $DTS$ set-up, ultrasound transducers can be placed in ports at 4 different latitudes ($10^\circ$, $-20^\circ$ and $\pm 40^\circ$) (see \citet{nataf08a} and \citet{brito11} for more details).
Purely radial shots provide radial profiles of the radial velocity.
Shooting at some angle away from the radial direction, we get profiles that record a combination of meridional and azimuthal velocities.
We observe that azimuthal velocities are one order of magnitude larger than the meridional velocities (we record a maximum azimuthal velocity of $1.8$ m/s for $f=\pm3$ Hz, whereas the maximum meridional velocity reaches $0.16$ m/s).
In a first approximation, the angular velocity in the fluid is thus simply the fluid velocity measured along the ultrasound beam \citep[{\it e.g.},][]{brito01}.

However, we note that the shape of the profiles depends slightly upon the direction of rotation of the inner sphere.
This is because, while the sign of the azimuthal velocity is reversed in this operation, the meridional flow keeps the same sign.
Therefore, I construct the sum and the difference of profiles shot for two opposite rotation frequencies $f$ and $-f$.
I thus obtain refined azimuthal velocity profiles from the difference, and meridional profiles from the sum.

The resulting spatial data coverage in the ($s$,$z$) upper quadrant is shown by the colored lines in figure \ref{fig:data} for the azimuthal velocity (left) and for the meridional velocity (right)
(we will see in section \ref{Field decomposition} that the meridional and azimuthal fields are decoupled under our approximations.)

In this study, I focus on time-averaged properties.
For a given rotation frequency $f$, we record a series of profiles over a time lapse $T$.
For each point of the profiles, the mean velocity is the time-averaged value, and the standard deviation of the mean $\tau_{mean}$ is given by $\tau_{mean} = \tau/\sqrt{T \, f}$, where $\tau$ is the root mean square deviation, and $T\, f$ is used as an estimate of the number of statistically independent samples.
I retain only one every 4 points along the ray because the original profiles were over-sampled. 

\subsubsection{magnetic field inside the fluid}
\label{sleeve}
The magnetic field is measured by Hall magnetometers placed on a narrow rectangular board inside a cylindrical stainless steel sleeve, which penetrates the fluid radially.
The sleeve can be placed at three different latitudes: $10^\circ$, $-20^\circ$ and $\pm40^\circ$.
The azimuthal field is measured at 6 different radii along the board, while the radial and orthoradial components are obtained at 2 radii.
We emplace only one sleeve at a time in order to limit the perturbation of the flow that it produces.
As for the Doppler profiles, the collection of data I invert was acquired in different campaigns, with $f$ ranging from $2.7$ to $4.0$ Hz.
All data are scaled by their actual $f$ before being combined. 
The resulting spatial data coverage in the ($s$,$z$) upper quadrant is shown by the colored symbols in figure \ref{fig:data} for the azimuthal field (left) and for the meridional field (right).

The mean azimuthal field is deduced from a time-average of the records for a given $f$ over a time-lapse of typically $300$ turns, and its standard deviation from the root mean square deviations divided by an estimate of the square root of independent samples as described in the previous section.
The zero reference is the average value measured when the inner sphere rotates very slowly (in order to average over small deviations away from a perfect dipole, which amount to about $1\%$ of the dipole field).

The induced meridional magnetic field is one order of magnitude smaller than the azimuthal field.
For $f=\pm3$ Hz, the maximum measured meridional field is $0.6$ mT, to be compared with $5.6$ mT for the azimuthal field.
The amplitude of the average field is comparable to that of its fluctuations.
It can also be contaminated by a projection of the much stronger azimuthal field if the alignment is not perfect.
As for the Doppler profiles, I take advantage of the differing symmetries of the azimuthal and meridional fields and correct for misalignment by making the values for the latter coincide when the sign of $f$ is reversed.
The misalignments are found to range between $1^\circ$ and $6^\circ$.
I then deduce the mean meridional field from a time-average of the records for a given $f$ over a time-lapse of typically $300$ turns.
The zero value remains difficult to establish because the value measured when the inner sphere is at rest varies by up to $\pm 0.2$ mT depending on the position in which it stopped, due to the deviations of the imposed field from a perfect dipole.
Therefore, I assume that the induced field varies linearly with $f$ for the small values of $f$ that I consider and obtain the value for $f=\pm3$ Hz by rescaling the difference between the average signal at $f=\pm3$ Hz and that at $f\simeq0.7$ Hz.
The induced meridional magnetic field is thus not very precisely measured in our experiment, which is reflected in its relatively large error bars.

\subsubsection{magnetic field at the surface}
\label{b_surf}

The three components of the magnetic field are measured at several latitudes from $-57^\circ$ to $57^\circ$ at the surface of the outer sphere.
The magnetometers are Giant Magneto Resistive (GMR) devices arranged on an in house designed electronic board (see \citet{schmitt12} for details).
The time-averaged azimuthal magnetic field vanishes at the surface of the fluid.
Therefore, the surface measurements only constrain the meridional magnetic field.
The actual spatial data coverage in the ($s$,$z$) upper quadrant is shown by the colored symbols at the surface of the sphere in figure \ref{fig:data} (right).
The maximum induced surface magnetic field we measure at $f=3$ Hz is $0.04$ mT, which is difficult to measure.
In particular, one must take into account the effect of temperature on the signals.
We correct for these, using platinum resistance thermometers installed on the GMR board, and calibration curves measured under controlled temperatures and imposed magnetic fields.
The high latitude radial probes are saturated by the dipole field and not used in the present analysis.
The zero value remains difficult to establish because we cannot use the value measured when the inner sphere is at rest, since it varies by up to $0.03$ mT depending on the position in which it stopped, due to the deviations of the imposed field from a perfect dipole.
Therefore, I assume that the induced field varies linearly with $f$ for the small values of $f$ that I consider and obtain the value for $f=\pm3$ Hz by rescaling the difference between the average signal at $f=\pm3$ Hz and that at $f\simeq1$ Hz.
The values obtained from the two hemispheres and for the two signs of rotation are consistent and are combined to yield my final data.

\subsubsection{electric potentials at the surface}
\label{ddp}

Electric currents are produced through the interaction of the fluid flow with the magnetic field.
Through Ohm's law, the electric field carries some information about these currents, together with the induction term $\boldsymbol{U} \times \boldsymbol{B}$.
The time-averaged potentials do not vary with azimuth, and thus relate only to the azimuthal flow.
I therefore include in my azimuthal inversion the differences between the electric potential measured at 5 different latitudes on the outer shell, as indicated by the squares in figure \ref{fig:data} (left).
The time-averaged values and their standard deviations are deduced from $640$s-long records.
The maximum electric potential difference between electrodes $10^\circ$ apart in latitude is $0.84$ mV for $f=\pm3$ Hz.

\subsubsection{torque on the inner sphere}
\label{torque}

We will see that the magnetic torque applied by the fluid on the inner sphere is directly related to the strength of the induced azimuthal magnetic field at its surface.
We expect the magnetic torque to dominate over the viscous torque on the inner sphere (whereas the opposite holds on the outer sphere).
I therefore include this data as a further constraint in the azimuthal inversion.
The torque is recorded for all runs, as given by the electronic drive of the motor.
We substract from the reading the mechanical torque estimated for the given rotation rate from runs without liquid sodium (see \citet{brito11} for further details).
The torque is equal to $-1.2 \pm 0.1$ Nm at $f = 3$ Hz, which translates into $-0.29 \pm 0.03$ in our dimensionless units (see \ref{adimensionalization}).

\section{Field decomposition}
\label{Field decomposition}

The problem that I consider here is purely kinematic: we do not solve for the dynamics but we want to relate the velocity field to the magnetic field it interacts with, using the induction equation:

\begin{equation}
\frac{\partial \boldsymbol{B}}{\partial t} =   \bnabla \times (\boldsymbol{U} \times \boldsymbol{B}) + \eta \Delta   \boldsymbol{B},
\label{eq:induction}
\end{equation}
where $\eta$ is the uniform magnetic diffusivity of the fluid.

Both the velocity field $\boldsymbol{U}$ and the magnetic field $\boldsymbol{B}$ are divergence-free and can therefore be decomposed into their poloidal and toroidal components.
I use spherical coordinates $(r, \theta, \varphi)$.
Since I only consider axisymmetric mean fields, which do not depend upon the azimuth $\varphi$, the decomposition can be further simplified into the following expressions ({\it e.g.}, \citet{roberts07}):

\begin{equation}
\boldsymbol{U} = \boldsymbol{u}_P +  U_{\varphi} \boldsymbol{e}_{\varphi} = \bnabla \times u \boldsymbol{e}_{\varphi} +  U_{\varphi} \boldsymbol{e}_{\varphi},
\label{eq:utot}
\end{equation}
where $u$ is the potential of the meridional flow and $U_{\varphi}$ is the azimuthal velocity field.
The magnetic field is given by:

\begin{equation}
\boldsymbol{B} = \boldsymbol{B}_P +  b_{\varphi} \boldsymbol{e}_{\varphi} = \boldsymbol{B}_d + \boldsymbol{b}_P +  b_{\varphi} \boldsymbol{e}_{\varphi}  =  \bnabla \times \left( A_d + a \right) \boldsymbol{e}_{\varphi} +  b_{\varphi} \boldsymbol{e}_{\varphi},
\label{eq:btot}
\end{equation}
where $A_d$ is the potential of the imposed dipolar field $\boldsymbol{B}_d$, $a$ the potential of the induced meridional magnetic field $\boldsymbol{b}_P$, and $b_{\varphi}$ the induced azimuthal magnetic field.

All fields are made dimensionless, using $\Omega^{-1}$ for a time scale and $r_o^*$ for a length scale.
The imposed dipolar magnetic field is scaled by its intensity $B_o$ at the equator of the outer sphere.
The expression of $\boldsymbol{B_d}$ is thus:

\begin{equation}
\boldsymbol{B}_d = \bnabla \times A_d \boldsymbol{e}_{\varphi} = \frac{1}{r^3} \left( 2 \; \cos \theta \; \boldsymbol{e}_r + \sin \theta \; \boldsymbol{e}_{\theta} \right).
\label{eq:dipole}
\end{equation}

The induced magnetic fields $\boldsymbol{b}_P$ and $b_\varphi$ are scaled by $\Rm B_o$, where $\Rm = \Omega (r_o^*)^2/ \eta$ is the magnetic Reynolds number, since we expect the induced field to vary almost linearly with $\Rm$ and with $B_o$ in the moderate magnetic Reynolds number regime that we consider.

Led by our experimental measurements, I further consider the following approximations: $U_\varphi >> u_P$ and $B_d >> b_\varphi >> b_P$.
In other words, the flow is predominantly azimuthal, and it induces a magnetic field that is also mostly azimuthal, and which remains small compared to the imposed dipolar field.

Under all these conditions, and assuming steady state, the induction equation decomposes into two independent scalar equations \citep[e.g.,][]{fearn88, barenghi91}:
The induction equation for the azimuthal field $b_{\varphi}$ yields:
\begin{equation}
s \boldsymbol{B}_d \cdot \bnabla \omega +\Delta_1 b_{\varphi} = 0,
\label{eq:az:induction}
\end{equation}
with $\Delta_1 = \nabla^2 -1/s^2$, and where we introduced the angular fluid velocity $\omega = U_\varphi/s$, with $s = r \sin \theta$ the cylindrical radius.
The induction equation of the meridional magnetic potential $a$ becomes:

\begin{equation}
\frac{\boldsymbol{u}_p}{s} \cdot \boldsymbol{\nabla}(s A_d) - \Delta_1 a = 0.
\label{eq:mer:induction}
\end{equation}

In classical mean-field dynamo theory, one requires mechanisms to convert large-scale poloidal magnetic field into large-scale toroidal magnetic field, and {\it vice versa}.
For rapidly rotating bodies, it is believed that azimuthal shear is an efficient way of achieving the former.
It corresponds to the induction term of equation \ref{eq:az:induction} and is therefore called the `$\omega$-effect'.
Our equation \ref{eq:mer:induction} for the meridional (poloidal) magnetic field only converts poloidal magnetic field (the imposed dipole) into poloidal magnetic field.
One usually resorts to the contribution of small-scale fluctuations to produce an `$\alpha$-effect', which converts toroidal magnetic field into poloidal magnetic field.
The objective of our study is to see how much can be explained by the large-scale flow alone, in order to measure the need for an $\alpha$-effect in our experiment.

\section{Azimuthal fields: forward problem}
\label{Azimuthal forward problem}

In this section, I want to express all the observables as functions of the azimuthal velocity field $U_\varphi$, which I will invert for.
$U_\varphi$ projects onto the associated Legendre functions $P_l^1$ of order $1$ as:

\begin{equation}
U_{\varphi}(r,\theta) = \sum_{odd \; l} U_l(r) P_l^1(\cos \theta),
\label{eq:az:Legendre_u}
\end{equation}
where only odd degrees $l$ are considered because $U_\varphi$ is symmetric with respect to the equator.

\subsection{induction equation}
\label{az:induction equation}

Similarly, $b_\varphi$ is written as:

\begin{equation}
b_{\varphi}(r,\theta) = \sum_{even \; n} b_n(r) P_n^1(\cos \theta),
\label{eq:az:Legendre_b}
\end{equation}
where the sum is over even degrees $n$ since we anticipate that $b_\varphi$ is equatorially antisymmetric (I use $n$ to denote even degrees and $l$ for odd degrees).
With this decomposition, the diffusion term of the induction equation (\ref{eq:az:induction}) becomes:
\begin{equation}
\Delta_1 b_{\varphi} = \sum_{even \; n} \left[ \frac{1}{r} \frac{d^2 (r b_n)}{d r^2} - \frac{n(n+1)}{r^2} b_n  \right] P_n^1(\cos \theta),
\label{eq:az:diffusion}
\end{equation}
while the induction term is expressed as:


\begin{equation}
\begin{split}
s  \; \boldsymbol{B}_d \cdot \boldsymbol{\nabla}\left( \frac{U_{\varphi}}{s}\right)
&= \frac{1}{r^4} \sum_{odd \; l} 
\left\{ 
U_l(r) \frac{d}{d \theta} \left( \sin \theta \; P_l^1 (\cos \theta)\right) \right. \\
& \left. \quad + 2 \left[ r \frac{d U_l(r)}{dr} - 2 U_l(r)\right] \cos \theta \, P_l^1 (\cos \theta)
\right\}.
\end{split}
\label{eq:az:induction_1}
\end{equation}
In order to keep all $\theta$ dependence in the associated Legendre polynomials only, we make use of the following relationships:
\begin{eqnarray}
\frac{d}{d \theta} \left( \sin \theta \; P_l^1 \right) &=& \frac{l(l+1)}{2l+1} \left\{ P_{l+1}^1 - P_{l-1}^1  \right\} \\
\cos \theta \; P_l^1 &=& \frac{1}{2l+1} \left\{ l \, P_{l+1}^1 + (l+1) \, P_{l-1}^1  \right\}.
\label{eq:recurrence}
\end{eqnarray}
Injecting into equation (\ref{eq:az:induction_1}), the induction term becomes:

\begin{equation}
\begin{split}
	s  \; \boldsymbol{B}_d \cdot \boldsymbol{\nabla}\left( \frac{U_{\varphi}}{s}\right) 
	&= \frac{1}{r^4} \sum_{odd \; l} \frac{1}{2l+1}
	\left\{
	\left[ (l-3) U_l + 2 r \frac{d U_l}{dr}\right] l \, P_{l+1}^1 \right. \\
&\left. \quad + \left[ -(l+4) U_l + 2 r \frac{d U_l}{dr}\right] (l+1) \, P_{l-1}^1
	\right\},
	\end{split}
	\label{eq:az:induction_2}
\end{equation}
which confirms that only even degrees $n$ contribute to the azimuthal magnetic field $b_\varphi$.
Equating equation \ref{eq:az:diffusion} and the opposite of equation \ref{eq:az:induction_2}, we obtain for each even degree $n$ a linear relationship between $b_n(r)$ and $U_{n \pm 1}(r)$.
\ref{parity and truncation} indicates the truncation degree to be considered for the various fields when $l_{max}$ is the truncation degree of the expansion of the azimuthal velocity field $U_\varphi$ in equation \ref{eq:az:Legendre_u}.

\subsection{electric field and potential}
\label{electric field and potential}

In order to compare with the differences in electric potential measured at the surface of the sphere, we
need to relate the electric potential $V$ to the azimuthal velocity $U_\varphi$.
The $E_\theta$ component of the electric field $\boldsymbol{E} = - \bnabla V$ can be decomposed as:
\begin{equation}
E_\theta(r,\theta) = \sum_{even \; n} e_n(r) P_n^1(\cos \theta),
\label{eq:Legendre_e}
\end{equation}
which we integrate to get:
\begin{equation}
V(r,\theta) = V(r) - \sum_{even \; n} r \; e_n(r) P_n^0(\cos \theta).
\label{eq:Legendre_V}
\end{equation}
We compute the radial function $V(r)$ by integrating $E_r$ along the $z$-axis.
The $E_\theta$ and $E_r$ components of the electric field are derived from Ohm's law:
\begin{equation}
\boldsymbol{E} = \boldsymbol{j} - \boldsymbol{U} \times \boldsymbol{B},
\label{eq:ohm}
\end{equation}
which yields:
\begin{equation}
E_\theta(r,\theta) = -\frac{\partial}{r\partial r} \left( r b_\varphi\right) -\frac{2}{r^3}\sum_{odd \; l} U_l(r) \left[ \frac{l+1}{2l+1}P_{l-1}^1+\frac{l}{2l+1}P_{l+1}^1\right].
\label{eq:E_theta}
\end{equation}
We get the $e_n(r)$ coefficients of equation \ref{eq:Legendre_e} by identification, using the expansion of $b_\varphi$ in $b_n$, which were related to $U_l$ in the previous paragraph.

\subsection{boundary conditions}
\label{az:boundary conditions}

We make use of the continuity of the magnetic field and of the tangential components of the electric field at the interface between the fluid and the solid shells.
At these interfaces, thin velocity boundary layers form in the fluid, which we will not be able to resolve from our Doppler profiles.
I therefore authorize azimuthal velocity discontinuities on the walls.
\ref{solid shells} details how we deal with these discontinuities, as well as those in the electrical conductivity between the three shells.
Note that electric currents are present in the solid shells and that the radial derivative of $b_\varphi$ is discontinuous.
In the end, we obtain a relation between the magnetic field and its radial derivative (on the fluid side) at the interface $r=r_i$ given by:
\begin{equation}
\left. \frac{d \, b_n}{dr} \right|_{r_i} = - \mathcal{P}_n \, b_n(r_i) - \mathcal{Q}_n,
\label{eq:az_bc}
\end{equation}
and similarly for $r=r_o$.
The expressions of $\mathcal{P}_n$ and $\mathcal{Q}_n$ are given by \ref{eq:derivative_4} for $r=r_i$.

\subsection{magnetic torque}
\label{magnetic torque}

The magnetic torque is the torque due to the Lorentz force acting on the copper shell.
We can compute it by integrating:
\begin{equation}
\Gamma_M = 4 \pi \int_{\hat{r}_i}^{r_i} dr \int_0^{\pi/2} d \theta \; r^3 \sin^2\theta \; \mathcal{F}_{\varphi}(r,\theta),
  \label{eq:magnetic_torque}
\end{equation}
where $\mathcal{F}_{\varphi}$ is the azimuthal component of the Lorentz force per unit volume given by:
\begin{equation}
\mathcal{F}_{\varphi}(r,\theta) = (\boldsymbol{j} \times \boldsymbol{B})_{\varphi} = \frac{1}{r^4} \left[ \frac{\partial}{\partial\theta}(\sin\theta \; b_{\varphi}) + 2 \cos\theta \frac{\partial}{\partial r}(r b_{\varphi}) \right].
\end{equation}
Since we could project $b_{\varphi}$ over the $P_l^1(\cos\theta)$, and because of the recursive relations given in equations \ref{eq:recurrence}, the azimuthal component of the Lorentz force also projects onto the 
$P_l^1(\cos\theta)$:

\begin{equation}
\mathcal{F}_{\varphi}(r,\theta) = \sum_{odd \; l} f_l(r) P_l^1(\cos \theta).
\label{eq:Legendre_Lorentz}
\end{equation}
Developing the expansion for $b_{\varphi}$, we obtain:
\begin{equation}
f_l(r) = \frac{1}{r^4} \left\{ \frac{l+2}{2l+3} \left[ (1-l) b_{l+1} + 2 r \frac{db_{l+1}}{dr}\right] \right.
\left. + \frac{l-1}{2l-1} \left[ (l+2) b_{l-1} + 2 r \frac{db_{l-1}}{dr} \right] \right\}.
\end{equation}


Getting back to the expression of the magnetic torque (equation \ref{eq:magnetic_torque}), we note that because of the orthogonality of the $P_l^m(\cos\theta)$, the only degree of the expansion of the Lorentz force that contributes is $l=1$, which only implies the degree 2 component of $b_{\varphi}$.
It is then straightforward to see that the magnetic torque is directly proportional to the value of  $b_2$ at the surface of the copper shell, and does not depend on its thickness.
We simply get:

\begin{equation}
\Gamma_M = - \frac{16 \pi}{5} b_2(r_i).
  \label{eq:magnetic_torque_2}
\end{equation}
Note that, more generally, the integral of the magnetic torque over the volume of a conductive shell reduces to a surface integral of the Maxwell tensor expressed as $B_r b_\varphi$.

\section{Azimuthal inversion}
\label{Azimuthal inversion}

\subsection{inversion set-up}
\label{Az:inversion set-up}

We now have all the expressions that relate our various measurements to the radial functions $U_l(r)$ of the different degrees $l$ of the expansion of the azimuthal velocity field.
In order to carry out the inversion, I define a regular radial grid of $Nr=137$ points between $r_i=74/210$ and $r_o=1$, on which the $U_l$ functions are discretized.
The model vector $M$ consists of $Nr \, (l_{max}+1)/2$ unknowns.

The data vector $D$ consists of a set of 5 azimuthal velocity profiles (with about 150 independent point measurements each) measured by ultrasound Doppler velocimetry, 16 azimuthal magnetic field measurements inside the fluid from 3 different latitudes, 4 electric potential differences at the surface, and the value of the torque on the inner sphere.
The spatial data coverage is shown in figure \ref{fig:data}.

\subsubsection{matrix formulation}
\label{Az:matrix formulation}

Radial derivatives are computed by finite difference.
The diffusion operator in equation \ref{eq:az:diffusion} is expressed in matrix form, taking into account the conditions at the boundaries expressed by equation \ref{eq:az_bc}.
Under these conditions, the inverse problem is linear: all measured quantities are linearly related to the model vector, which consists of the $U_l(r_k)$ values.
Note that the contribution of the spinning inner sphere is treated separately, since there is no need to invert for this known (imposed) velocity.

The best fitting model $M$ is obtained with the classical generalized least square inverse method \citep{tarantola82}:

\begin{equation}
M = C_{pp} G^T \left(C_{dd} + G C_{pp} G^T \right)^{-1} D,
\label{eq:inverse}
\end{equation}
where $G$ is the matrix from which the predicted data are computed by $D_{fit}=GM$.
$C_{dd}$ is the covariance matrix of the data, and $C_{pp}$ the {\it a priori} covariance matrix of the model parameters, which I describe below.

\subsubsection{covariance matrices}
\label{Az:covariance matrices}

The covariance matrix of the data is taken diagonal.
The diagonal terms are the square of the standard deviations presented in the data processing section \ref{Data processing}.
I use the {\it a priori} covariance matrix of the model to control the smoothness of the model in the radial direction, while the smoothness in the latitudinal direction is controlled by the truncation degree $l_{max}$ of the Legendre expansion.
Within an $l$-block of the {\it a priori} covariance matrix of the parameters, the elements are taken as:
\begin{equation}
C_{pp}(r_a,r_b) = \left[ \frac{\tau_0}{l} \right]^2 \exp \left[ {- \left( \frac{r_a - r_b}{\delta} \right) ^2} \right],
\label{eq:Cpp}
\end{equation}
where $\delta$ is a smoothing parameter and $\tau_0/l$ is a damping parameter.

In order to limit the velocity jump at the inner and outer spheres, where the Doppler profiles cannot resolve the boundary layers, we add fake profiles with zero velocity and a large error bar ($\tau=0.1$).

\subsection{inversion results}
\label{Az:inversion results}

I present the results of a simultaneous inversion of the velocity profiles, the induced magnetic field inside the fluid, the electric potential differences at the surface, and the torque on the inner sphere.
The data coverage is shown in figure \ref{fig:data}.
The Legendre expansion of the azimuthal velocity is carried up to $l_{max}=7$ ($l$ odd), and the parameters of the a priori covariance matrix in expression \ref{eq:Cpp} are chosen as $\delta=0.2$ and $\tau_0=0.01$.

\subsubsection{radial profiles of the modes}
\label{Az:modes}

\begin{figure}
	\centerline{\includegraphics[width=17cm]{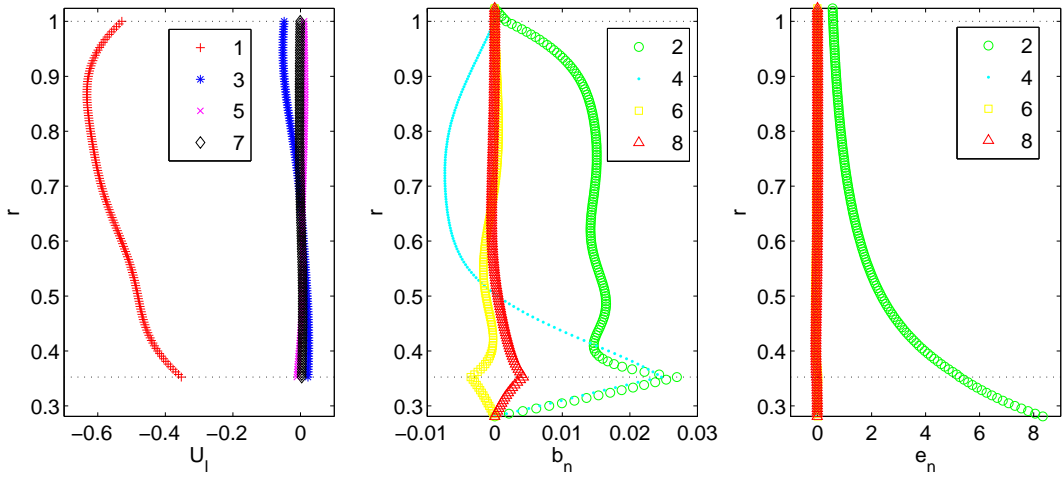}}
	\caption{Azimuthal inversion: radial profiles of the modes of different harmonic degree.
From left to right: radial functions of the azimuthal velocity $U_l(r)$, the azimuthal magnetic field $b_n(r)$ and the orthoradial electric field $e_n(r)$.
All fields are dimensionless as given in \ref{adimensionalization}.
The radius axis extends from $\hat{r}_i$ to $\hat{r}_o$.
horizontal dotted lines indicate the fluid/solid interfaces at $r_i$ and $r_o$.}
	\label{fig:az:modes}
\end{figure}

The radial profiles of the Legendre modes are shown in figure \ref{fig:az:modes} for the coefficients of the azimuthal velocity $U_l$, the azimuthal magnetic field $b_n$ and the $\theta-$component of the electric field $e_n$.
The $l=1$ mode dominates the velocity field.
The magnetic field modes show that the induction at the inner sphere boundary is strong, but that the shear inside the fluid produces large dipole ($n=2$) and octupole ($n=4$) field as well.
The electric field mode is almost entirely $n=2$ and does not drop to zero at the outer surface ($r=\hat{r}_o$).

\subsubsection{maps of the fields}
\label{Az:maps}

\begin{figure}
	\centerline{\includegraphics[width=15cm]{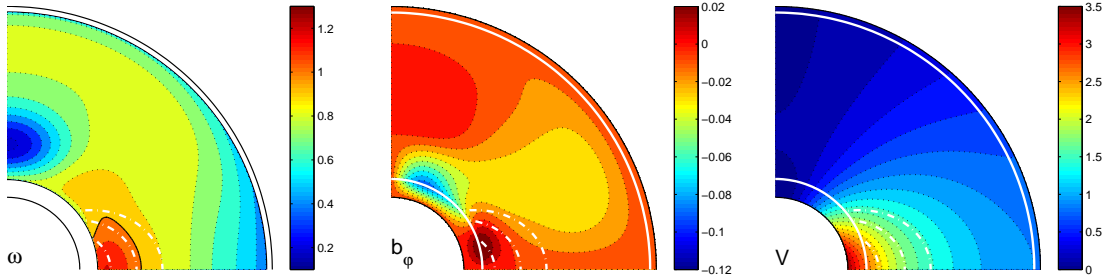}}
	\caption{Contour maps of the reconstructed angular velocity of the fluid (left), induced azimuthal magnetic field (center), and electric potential (right), from the inversion of $DTS$ measurements.
The amplitudes are given by the color bars.
All fields are dimensionless as given in \ref{adimensionalization}.}
	\label{fig:az:maps}
\end{figure}

Figure \ref{fig:az:maps} shows contour maps (in the northern meridional quadrant) of the angular velocity of the fluid $\omega$, the azimuthal magnetic field $b_\varphi$, and the electric potential $V$.
All quantities are dimensionless, as given in \ref{adimensionalization}.

The map of angular velocity illustrates some important properties of the flow already discussed by \citet{brito11}.
Note the region near the equator of the inner sphere where the angular velocity of the fluid gets larger ($ 1 < \omega < 1.3$) than the angular velocity of the inner sphere.
In that region, the fluid obeys Ferraro's law of isorotation \citep{ferraro37}: the angular velocity is constant along magnetic field lines (white dashed lines).
Further away from the inner sphere, the mean flow is geostrophic: the velocity does not vary along the rotation axis (vertical axis).
Also note that the magnetic coupling between the inner copper shell and the fluid is very efficient and entrains the fluid at large angular velocities throughout the fluid volume.
There is a region of marked weak velocity at the pole of the inner sphere, for which we have no explanation yet.

The map of the induced azimuthal field illustrates that the field is mainly produced in two distinct regions: at the surface of the inner sphere, with alternating polarities, and in the large geostrophic shear zone.
Note that the amplitudes range from $-0.08$ to $0.02$ in the $\Rm B_0$ units we chose for $b_\varphi$ and that $\Rm \simeq 9.4$ for $f=3$ Hz.

The electric potential map appears dominated by the geometry of the imposed dipole field.

\subsubsection{fits to the data}
\label{Az:fits}

\begin{figure}
	\centerline{\includegraphics[width=17cm]{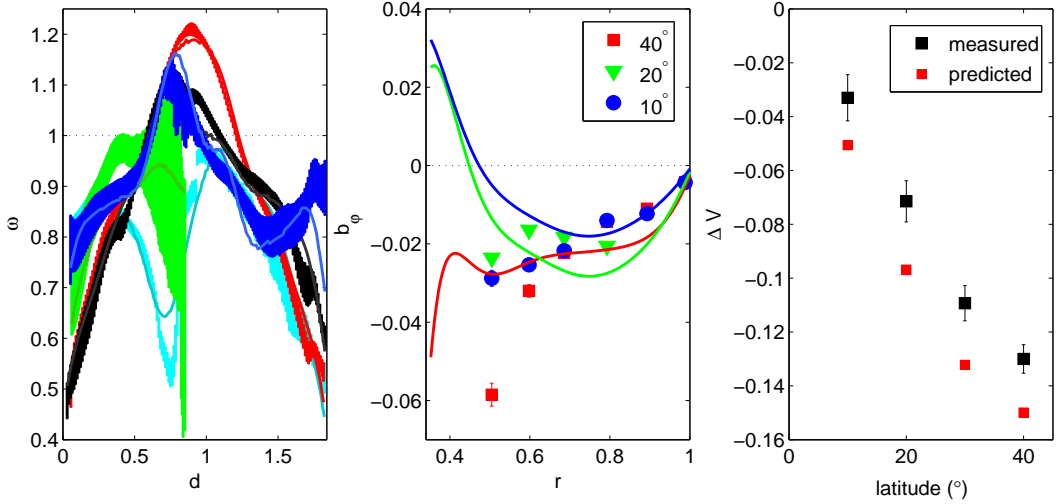}}
	\caption{Azimuthal inversion: fits to the data.
These plots compare the measured data, with their error bars, to the predictions of our inversion model for the azimuthal fields.
We use the same colors and symbols as in figure \ref{fig:data} (left), which gives the spatial distribution of the measurements.
All data are dimensionless as given in \ref{adimensionalization}.
{\it Left:} angular velocity from Doppler velocimetry, as a function of the distance $d$ along the ray of the ultrasound beam. {\it Center:} azimuthal magnetic field measured inside the fluid at different radii for 3 latitudes. {\it Right:} differences in electric potential between electrodes $10^\circ$ degrees apart at the surface of the outer shell.}
	\label{fig:az:fits}
\end{figure}

Let's now see how well the reconstructed azimuthal velocity field fits the observations under the $\omega$-effect approximation. Figure \ref{fig:az:fits} compares the predictions of our model with the velocity measurements along the ultrasound Doppler profiles, the azimuthal magnetic field inside the fluid, and the electric potential differences at the surface. In all cases, we compute the normalized misfit defined as:

\begin{equation}
misfit = \sqrt{\frac{1}{N} \sum^N{\left( \frac{d_{pred} - d_{obs}}{\tau_{obs}}\right)^2}},
\label{eq:misfit}
\end{equation}
where $d_{obs}$ are the observed data with their error $\tau_{obs}$, and $d_{pred}$ the model prediction.

The ultrasound Doppler profiles are well fit by our simple velocity model. The normalized misfit is $1.9$. The fact that it is larger than one indicates that representation errors are present. Some of our assumptions (such as symmetry with respect to the equator) are partially violated. We observe for example that slightly different velocities are measured at points where Doppler profiles intersect in the ($s$,$z$) plane.

The predicted induced azimuthal magnetic field under the $\omega$-effect approximation ({\it i.e.,} taking into account the induction from the mean flow alone) falls in the right range, but deviates strongly from the observations at all latitudes for $r < 0.6$.
The normalized misfit is $8$.

With a normalized misfit of $3.2$, the differences in electric potential at the surface are not perfectly fit, but display the proper trend.

The predicted magnetic torque is $-0.27$, slightly smaller than the observed value of $-0.29 \pm 0.03$.

I find it quite remarkable that the simple velocity model I invert for provides a reasonable fit to observations as diverse as induced magnetic field, electric potential and torque, when plugged into the mean flow induction equation.
By playing with the {\it a priori} covariance matrix, it is possible to improve the fit to the observations, at the expense of added complexity in the velocity field.
However, the velocity field is very strongly constrained by our direct measurements of azimuthal velocity along criss-crossing profiles inside the fluid, and I have not been able to find models that could significantly improve the fit of the observed induced magnetic field $b_\varphi$ inside the fluid.

\section{Meridional fields: forward problem}
\label{Meridional forward problem}

We now consider the poloidal (meridional) part of the velocity field.
According to equation \ref{eq:utot}, the poloidal velocity field is expressed as:

\begin{equation}
\boldsymbol{u}_p = \boldsymbol{\nabla} \times u \boldsymbol{e}_{\varphi}.
\label{eq:u_p}
\end{equation}
The direction of rotation of the meridional flow changes sign at the equator, which implies that the potential $u$ projects on associated Legendre polynomials with even degree $n$:
\begin{equation}
u(r,\theta) = \sum_{even \; n} u_n(r) P_n^1(\cos \theta).
\label{eq:mer:Legendre_u}
\end{equation}
We derive the expressions of the $u_r$ and $u_\theta$ components, which we use to relate to the velocity profiles measured by ultrasound Doppler velocimetry:

\begin{eqnarray}
u_r(r,\theta) &=& \sum_{even \; n} \frac{u_n(r)}{r} \frac{1}{\sin \theta} \frac{d}{d \theta} \left( \sin \theta \; P_n^1 (\cos \theta)\right) \\
u_{\theta}(r,\theta) &=& - \sum_{even \; n} \left( \frac{u_n(r)}{r} + \frac{d u_n(r)}{d r} \right) \; P_n^1 (\cos \theta).
\label{eq:u_r_t}
\end{eqnarray}

\subsection{induction equation}
\label{Mer:induction equation}

Our next step is to derive a linear relation between the potential $a$ of the magnetic field and the potential $u$ of the flow from the induction equation \ref{eq:mer:induction}, which I recall here:
\begin{equation}
\frac{\boldsymbol{u}_p}{s} \cdot \boldsymbol{\nabla}(s A_d) - \Delta_1 a = 0.
\label{eq:mer:induction_bis}
\end{equation}
The potential $a$ projects onto the associated Legendre polynomials:
\begin{equation}
a(r,\theta) = \sum_{odd \; l} a_l(r) P_l^1(\cos \theta).
\label{eq:mer:Legendre_b}
\end{equation}
where the sum is over odd degrees $l$ since we anticipate that $a$ is symmetric with respect to the equator.
With this decomposition, the diffusion term of the induction equation \ref{eq:mer:induction_bis} becomes:
\begin{equation}
\Delta_1 a = \sum_{odd \; l} \left[ \frac{1}{r} \frac{d^2 (r a_l)}{d r^2} - \frac{l(l+1)}{r^2} a_l  \right] P_l^1(\cos \theta),
\label{eq:mer:diffusion}
\end{equation}
while the induction term becomes:
\begin{equation}
\frac{\boldsymbol{u}_p}{s} \cdot \boldsymbol{\nabla}(s A_d) = \frac{1}{r^3} \left( - u_r \sin \theta + 2 u_{\theta} \cos \theta \right),
\label{eq:mer:induction_1}
\end{equation}
since the potential $A_d$ of the dipolar field is $(\sin \theta) / r^2$.
Injecting the expressions for $u_r$ and $u_{\theta}$ (equations \ref{eq:u_r_t}), one gets the following expression:
\begin{equation}
\frac{\boldsymbol{u}_p}{s} \cdot \boldsymbol{\nabla}(s A_d) = - \frac{1}{r^4} \sum_{even \; n} 
\left\{ 
u_n \frac{d}{d \theta} \left( \sin \theta \; P_n^1 (\cos \theta)\right) \right.
\left. + 2 \left( u_n + r \frac{d u_n}{dr}\right) \cos \theta \, P_n^1 (\cos \theta)
\right\}.
\label{eq:mer:induction_2}
\end{equation}
%
%
Injecting the recursive relations \ref{eq:recurrence} into the previous equation, we finally obtain:
\begin{equation}
\begin{split}
\frac{\boldsymbol{u}_p}{s} \cdot \boldsymbol{\nabla}(s A_d) &= - \frac{1}{r^4} \sum_{even \; n} \frac{1}{2n+1}
\left\{
\left[ (3+n) u_n + 2 r \frac{d u_n}{dr}\right] n \, P_{n+1}^1\right. \\
& \quad \left.+ \left[ (2-n) u_n + 2 r \frac{d u_n}{dr}\right] (n+1) \, P_{n-1}^1,
\right\}.
\end{split}
\label{eq:mer:induction_3}
\end{equation}
%
%
which confirms that only odd degrees $l$ contribute to the potential $a$ of the meridional magnetic field $\boldsymbol{b}_P$.
Equating equation \ref{eq:mer:diffusion} and equation \ref{eq:mer:induction_3}, we obtain for each odd degree $l$ a linear relationship between $a_l(r)$ and $u_{l \pm 1}(r)$.
\ref{parity and truncation} recalls the truncation degree to be considered for the various fields when $n_{max}$ is the truncation degree of the expansion of the meridional velocity field potential $u$ in equation \ref{eq:mer:Legendre_u}.

\subsection{boundary conditions}

We impose non-penetration of the fluid at the interface with the solid shells.
However, as for the azimuthal velocity, I allow for a discontinuity of the tangential velocity $u_\theta$ since we cannot resolve the very thin boundary layers.
The continuity of the magnetic field implies that both $a$ and its radial derivative are continuous.
Only azimuthal currents contribute to the meridional magnetic field.
Thus there is no electric current in the solid shells since both the azimuthal electric field and the $\boldsymbol{U} \times \boldsymbol{B}$ contribution to Ohm's law (equation \ref{eq:ohm}) are null.
It means that the meridional field matches a potential field on both sides of the fluid shell, yielding a simple relation between $a$ and its derivative: 
\begin{eqnarray}
\left. \frac{d a_l}{dr} \right|_{r_i} &=& l \; a_l(r_i) \\
\left. \frac{d a_l}{dr} \right|_{r_o} &=& -(l+1) \; a_l(r_o).
\label{eq:mer_bc}
\end{eqnarray}

\section{Meridional inversion}
\label{Meridional inversion}

\subsection{inversion set-up}
\label{Mer:inversion set-up}

We now have all the expressions that relate our various measurements to the radial functions $u_n(r)$ of the different degrees $n$ of the expansion of the meridional velocity potential.
In order to carry out the inversion, I use the same regular radial grid of $Nr=137$ points between $r_i=74/210$ and $r_o=1$, on which the $u_n$ functions are discretized.
The model vector $M$ consists of $Nr \, n_{max}/2$ unknowns.

The data vector $D$ consists of a set of 8 Doppler profiles (radial profiles of the radial velocity at 3 latitudes, one meridional profile, and meridional velocities retrieved from 4 off-meridional profiles by summing azimuthal profiles for opposite rotation direction), measurements of $b_r$ and $b_\theta$ inside the fluid at 3 latitudes, and measurements of $b_r$ and $b_\theta$ at the surface from the equator up to $57^\circ$ in latitude.
The spatial data coverage is shown in figure \ref{fig:data}.

\subsubsection{matrix formulation}
\label{Mer:matrix formulation}

We follow the same approach as for the azimuthal inversion.
Radial derivatives are computed by finite difference.
The diffusion operator in equation \ref{eq:mer:diffusion} is expressed in matrix form, taking into account the conditions at the boundaries expressed by equation \ref{eq:mer_bc}.
Under these conditions, the inverse problem is linear: all measured quantities are linearly related to the model vector, which consists of the $u_n(r_k)$ values.
The best fitting model $M$ is obtained using the classical generalized least square inverse given by equation \ref {eq:inverse}.

$C_{dd}$ is the covariance matrix of the data, and $C_{pp}$ the {\it a priori} covariance matrix of the model parameters, which I describe below.

\subsubsection{covariance matrices}
\label{Mer:covariance matrices}
The covariance matrix of the data is taken diagonal.
The diagonal terms are the square of the standard deviations presented in the data processing section \ref{Data processing}.
The {\it a priori} covariance matrix of the model parameters has the same expression \ref{eq:Cpp} as for the azimuthal inversion. 
In order to impose non-penetration of the fluid at the inner and outer shells, I force $u_n(r_i)=u_n(r_o)=0$ for all degrees $n$.

\subsection{inversion results}
\label{Mer:inversion results}

I present the results of a simultaneous inversion of the velocity profiles and the induced magnetic field inside the fluid and at the surface.
The data coverage is shown in figure \ref{fig:data}.
The Legendre expansion of the meridional velocity potential is carried up to $n_{max}=8$ ($n$ even), and the parameters of the a priori covariance matrix in expression \ref{eq:Cpp} are chosen as $\delta=0.15$ and $\tau_0=0.001$.

\subsubsection{radial profiles of the modes}
\label{Mer:modes}

\begin{figure}
	\centerline{\includegraphics[width=16cm]{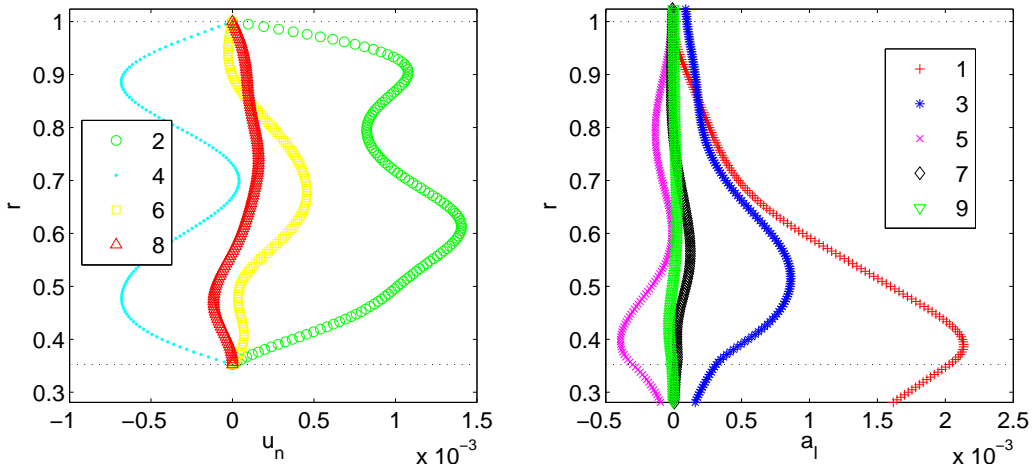}}
	\caption{Meridional inversion: radial profiles of the modes of different harmonic degrees.
From left to right: radial functions of the meridional velocity potential $u_n(r)$, and of the meridional magnetic potential $a_l(r)$.
All fields are dimensionless as given in \ref{adimensionalization}.
The radius axis extends from $\hat{r}_i$ to $\hat{r}_o$.
horizontal dotted lines indicate the fluid/solid interfaces at $r_i$ and $r_o$.	}
	\label{fig:mer:modes}
\end{figure}

The radial profiles of the Legendre modes are shown in figure \ref{fig:mer:modes} for the coefficients of the meridional potentials of velocity ($u_n$) and magnetic field ($a_l$) .
All quantities are dimensionless, as given in \ref{adimensionalization}.
Note that the velocity modes show more structure in radius than the azimuthal velocity modes.

Note that the induced dipole field ($l=1$) dominates near the inner sphere, but vanishes at the surface as it should \citep{spence06}.

\subsubsection{maps of the fields}
\label{Mer:maps}

\begin{figure}
	\centerline{\includegraphics[width=12cm]{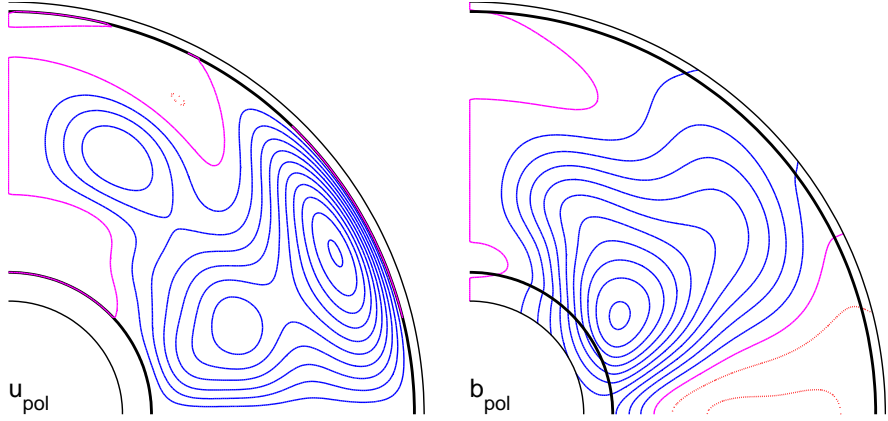}}
	\caption{Stream lines of the meridional flow (left) and field lines of the induced meridional magnetic field (right) reconstructed from the inversion of $DTS$ measurements.
The flow is poleward beneath the outer shell.
The values of the stream function range from $-2.0 \; 10^{-3}$ to $0.2 \; 10^{-3}$.
Those for the magnetic field from $-1.0 \; 10^{-3}$ to $0.2 \; 10^{-3}$.
The blue solid lines indicate anti-clock-wise circulation, and the pink dotted lines clok-wise circulation. The zero line is magenta.
}
	\label{fig:mer:maps}
\end{figure}

I show the streamlines of the meridional flow, and the field lines of the induced poloidal magnetic field in figure \ref{fig:mer:maps}.

The meridional circulation is poleward beneath the outer boundary, where the lines are squeezed, indicating large orthoradial velocities.
Our solution is very similar to that of \citet{brito11} who inverted the velocity data alone, using a sine decomposition in radius.

Not surprisingly, the induced magnetic field is largest near the inner sphere, where the imposed dipole is strongest.
The field lines are clockwise around the equator.
Note that the induced field remains very small compared to the imposed dipole field, in agreement with the approximation used in our forward problem (see section \ref{Field decomposition}
).
Indeed, I have plotted the field lines of the total field in figure \ref{fig:set-up} (dashed lines), but they hardly depart from those of the dipole.

\subsubsection{fits to the data}
\label{Mer:fits}

\begin{figure}
	\centerline{\includegraphics[width=17cm]{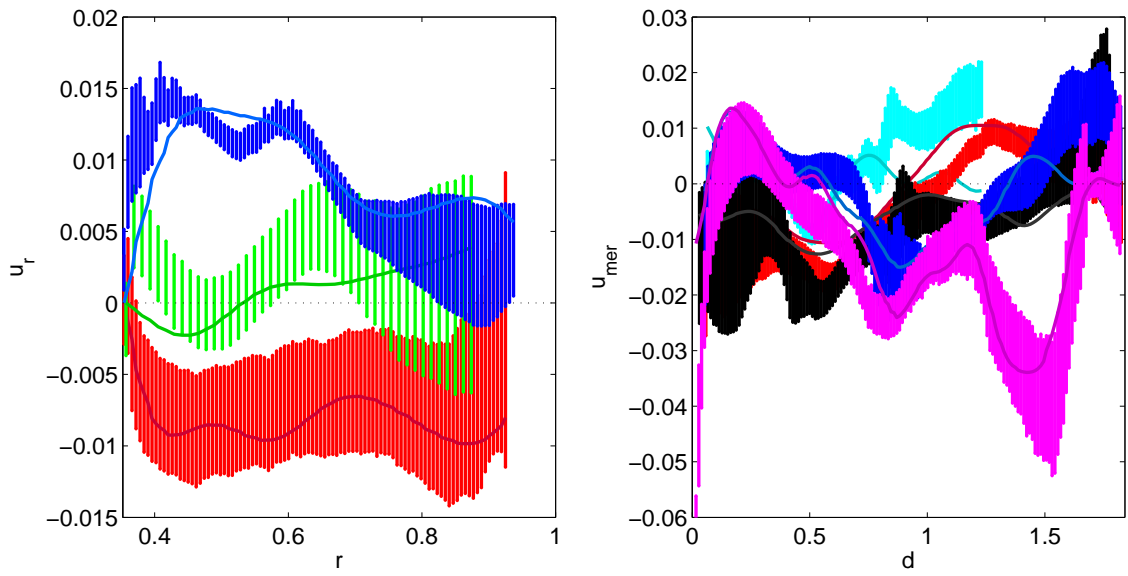}}
	\caption{Meridional inversion: fits to the velocity data.
These plots compare the measured data, with their error bars, to the predictions of our inversion model for the meridional velocity field.
We use the same colors as in figure \ref{fig:data} (right), which gives the spatial distribution of the measurements.
All data are dimensionless as given in \ref{adimensionalization}.
{\it Left:} radial velocity from the radial Doppler velocimetry profiles at three latitudes, as a function of radius.
{\it Right:} meridional velocity projected on the ultrasound beam as a function of the distance $d$ along the beam for the non-radial profiles.}
	\label{fig:mer:fits_1}
\end{figure}

Let's now see how well the reconstructed meridional velocity field fits the observations. Figure \ref{fig:mer:fits_1} compares the predictions of our model with the measured radial velocity profiles and the other meridional velocity profiles, all obtained by Doppler acoustics.
The fit looks rather good, with an overall normalized misfit of $1.2$.
The more complex radial structure, as compared to the azimuthal case, seems required by the measured profiles.

\begin{figure}
	\centerline{\includegraphics[width=17cm]{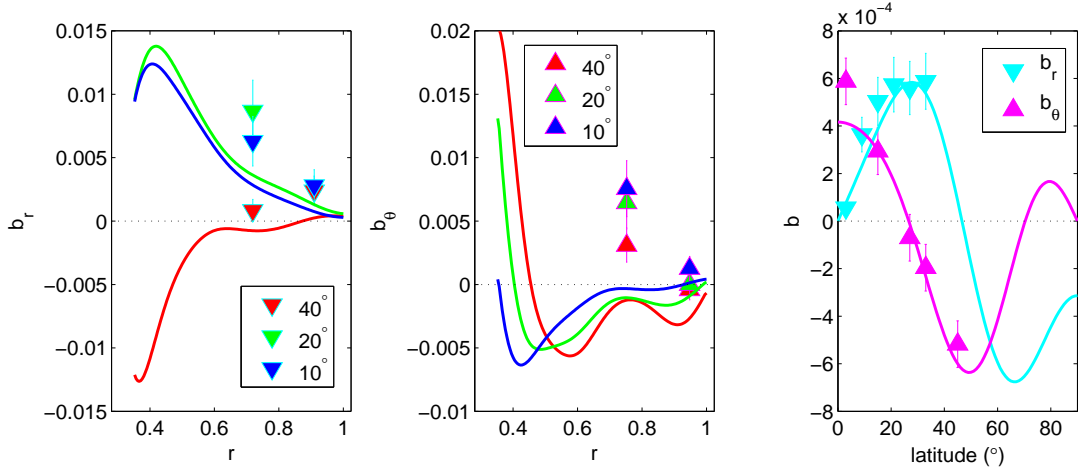}}
	\caption{Meridional inversion: fits to the magnetic data. 
These plots compare the measured data, with their error bars, to the predictions of our inversion model for the induced meridional magnetic field.
We use the same colors and symbols as in figure \ref{fig:data} (left), which gives the spatial distribution of the measurements.
All data are dimensionless as given in \ref{adimensionalization}.
{\it Left:} radial magnetic field measured inside the fluid at different radii for 3 latitudes. {\it Center:} orthoradial magnetic field measured inside the fluid at different radii for 3 latitudes. {\it Right:} radial and orthoradial components of the induced magnetic field measured at the surface of the outer shell.}
	\label{fig:mer:fits_2}
\end{figure}

Figure \ref{fig:mer:fits_2} compares the observations to the predicted induced magnetic field, under the approximation that it all results from induction by the mean meridional flow alone,
The $b_r$ and $b_\theta$ components inside the fluid are given in the first two panels.
The model predicts reasonable amplitudes, which are in the range $\pm 0.01$ in the $\Rm B_0$ units we chose for the induced field.
Unfortunately, our data coverage is rather sparse, and does not extend deep into the fluid.
The overall normalized misift is $2.4$.
The trends for $b_r$ are compatible with the data, while there is a clear disagreement for $b_\theta$.

The right panel of figure \ref{fig:mer:fits_2} shows the predicted magnetic field at the surface of the outer shell.
It fits rather nicely the data we measure, with a normalized misfit of $1.1$.

As in the azimuthal case, I find it rather remarkable that the velocity model, which is strongly constrained by our Doppler profiles, predicts rather well our magnetic observations.
It turns out that the magnetic field at the surface is quite sensitive to subtle changes in the velocity field.
In the case I present, I over-emphasized the importance of the surface measurements to obtain a good fit.
The slight modifications it brought to the velocity potential did not degrade the fit to the Doppler profiles.
In contrast, I was not able to provide a better fit to the $b_\theta$ field measured inside the fluid without introducing velocity features in contradiction with the velocity measurements.
Remember however that all meridional observables are at least one order of magnitude less than their azimuthal counterpart, and that some biases cannot be totally excluded.

\section{Discussion}
\label{Discussion}

In this article, I have exposed the formalism with which one can model the induced azimuthal magnetic field from the azimuthal velocity field (the $\omega$-effect), as well as other related measurable quantities (electric potentials and magnetic torque), taking into account realistic conductivity and velocity boundary conditions (section \ref{Azimuthal forward problem}).
I have written the inverse problem and inverted simultaneously ultrasound Doppler velocity profiles and magnetic observations from the $DTS$ experiment, in order to recover the mean azimuthal flow (section \ref{Azimuthal inversion}).
I have carried out a similar procedure for the meridional flow and the magnetic field it induces (sections \ref{Meridional forward problem} and \ref{Meridional inversion}).

The results are rather remarkable, in that I obtain a full coherent solution for the mean velocity and magnetic fields inside the fluid (figures \ref{fig:az:maps} and \ref{fig:mer:maps}), which can be used to evaluate in a quantitative fashion the distribution of energies and dissipation, the force balance, and so on.
In a sense, we have performed a numerical simulation of a magnetohydrodynamic flow at high Reynolds number ($\Rey \simeq 10^6$), using the experiment as a Navier-Stokes solver, and working out the induction equation numerically.

As an illustration, I report in table \ref{tab:energies} the magnetic and kinetic energies of the azimuthal and meridional flows.
They were obtained by integration of our solutions over the fluid shell (with the addition of the inner and outer solid shells for the magnetic field).
All energies are scaled by  $\rho \Omega^2 (r_o^*)^5$, yielding a value of order 1 for the kinetic energy of the azimuthal flow.
Note that kinetic energies $E_K^{az}$ and $e_K^{mer}$ are larger their magnetic counterparts $e_M^{az}$ and $e_M^{mer}$.
The magnetic energy of the imposed dipole field, whose energy, given by $E_M^{dipole} = 51/\Rm^2$, is in fact independent of the spin rate of the inner sphere.
For the spin rate $f=3$ Hz considered in this article ($\Rm \simeq 9.4$), the energy of the azimuthal flow $E_K^{az}$ is comparable to that of the imposed magnetic field.

\begin{table}[h] 
 	\begin{center}
		\begin{tabular}{ccccccc}
			$E_M^{dipole}$ & $e_M^{az}$ & $e_M^{mer}$ & $E_K^{az}$ & $e_K^{mer}$ & \multicolumn{1}{|c}{$\tilde{e}_M$} & $\tilde{e}_K$ \\
			\hline
			$51/\Rm^2$ & $1.9 \, 10^{-4}$ & $2.3 \, 10^{-5}$ & $0.45$ & $4 \, 10^{-4}$ & \multicolumn{1}{|c}{$\sim 2 \, 10^{-6}$} & $\sim 4 \, 10^{-3}$ \\
		\end{tabular}
		\caption{\label{tab:energies} Energies of the imposed dipolar field, the azimuthal magnetic field and the meridional magnetic field, and of the azimuthal and meridional flows, all in units of $\rho \Omega^2 (r_o^*)^5 = \rho \eta^2 r_o^* \Rm^2$.
		The table also lists estimates of the energies of the magnetic and velocity fluctuations $\tilde{e}_M$ and $\tilde{e}_K$ deduced from \citet{figueroa12} in the same units.}
	\end{center}
\end{table}

The mean flow and magnetic field in the $DTS$ experiment have been examined by \citet{brito11}, who show that at first order both the azimuthal and meridional velocities and the induced azimuthal magnetic field increase linearly with the rotation frequency $f$ of the inner sphere (or equivalently with the magnetic Reynolds number $\Rm$) for $f$ up to $30$ Hz.
The induced meridional magnetic field evolves in a more complex fashion, but I will assume here that it remains linear in $\Rm$ as well.
As a consequence, the fields we have inverted for at $f=3$ Hz give a rough first order image of the fields at higher rotation frequency.

Mean-field theory indicates that correlated fluctuations of the velocity and magnetic fields can contribute to the axisymmetric mean fields we are dealing with.
Fluctuations in the $DTS$-experiment are described in \citet{schmitt08, schmitt12, figueroa12}.
\citet{figueroa12} show that velocity and magnetic fluctuations also increase linearly with $\Rm$ at first order, and give an estimate for the energy of the fluctuations, which is listed in table \ref{tab:energies} for comparison.

I will use my inversion results and these various estimates to evaluate the contribution of the terms that have been left out of the induction equation under my simplifying assumptions.
The complete induction equation for the axisymmetric steady state, which replaces equations \ref{eq:az:induction} and \ref{eq:mer:induction},  can be written as:

\begin{equation}
\left \{ \begin{array}{rl}
\Delta_1 b_{\varphi} &=  - s \boldsymbol{B}_d \cdot \bnabla \omega + \Rm \left\{ - s \boldsymbol{b}_p \cdot \bnabla \omega + s \boldsymbol{u}_p \cdot \bnabla \left(\frac{b_{\varphi}}{s}  \right) - \left[\bnabla \times \langle \tilde{\boldsymbol{u}} \times \tilde{\boldsymbol{b}} \rangle \right]_{\varphi} \right\} \\
\Delta_1 a &=  \frac{\boldsymbol{u}_p}{s} \cdot \boldsymbol{\nabla}(s A_d) + \Rm \left\{ \frac{\boldsymbol{u}_p}{s} \cdot \boldsymbol{\nabla}(s a) \, - \langle\tilde{\boldsymbol{u}} \times \tilde{\boldsymbol{b}} \rangle_\varphi \right\}.
\end{array} \right.
\label{eq:alpha}
\end{equation}
All symbols have the same meaning and adimensionalization as in the rest of this article, yielding the $\Rm$-dependence of the additional terms.
Let us now check that these additional terms are negligible indeed at the rotation rate we considered ($f=3$ Hz, which yields $\Rm=9.4$), and evaluate their influence at the highest rotation rate we achieve ($f=30$ Hz, yielding $\Rm=94$). 
Before doing so, we examine the various additional terms inside the braces of equation \ref{eq:alpha}.

\begin{figure}
	\centerline{\includegraphics[width=6cm]{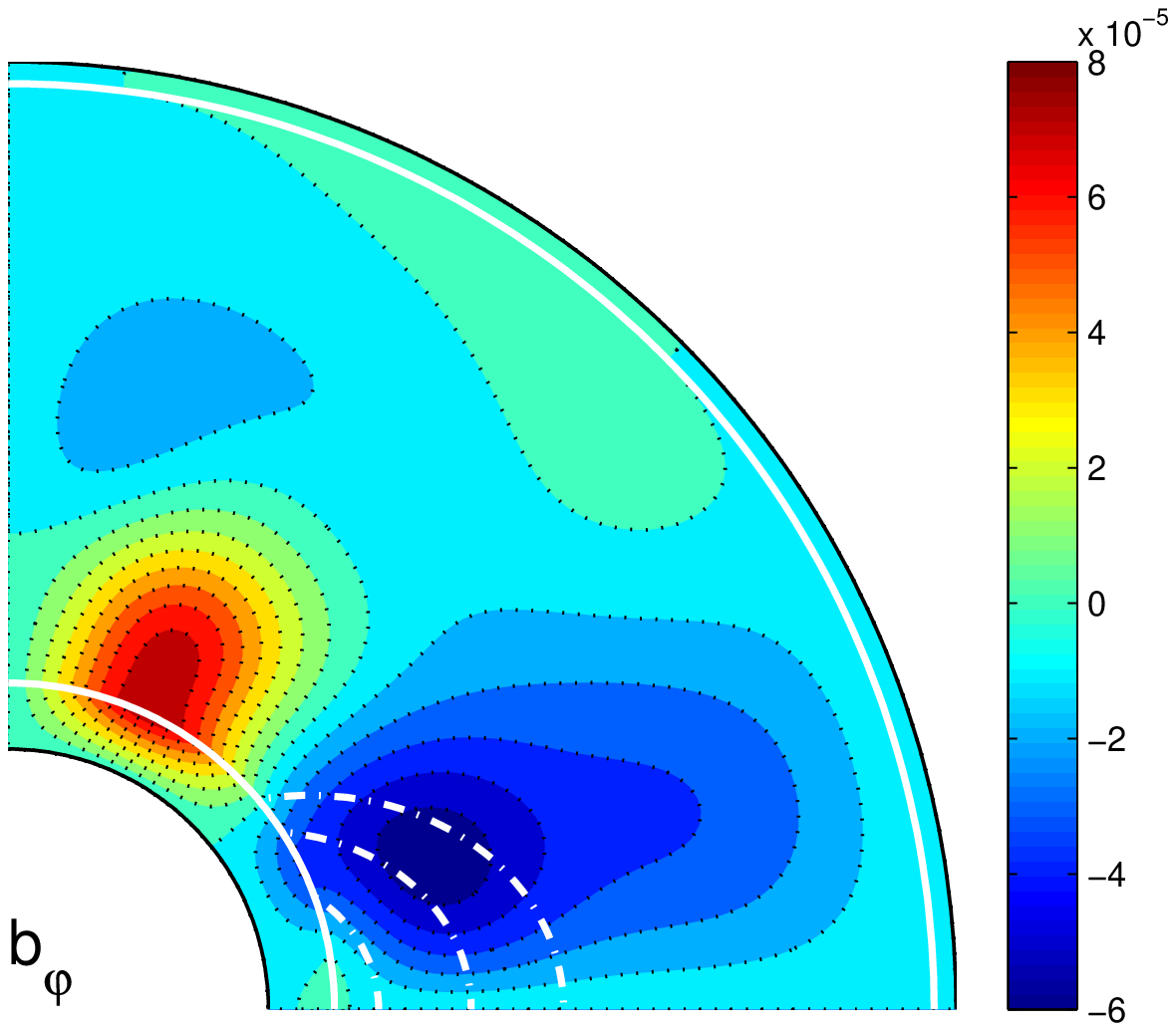} \hspace{-0.7cm} \includegraphics[width=6cm]{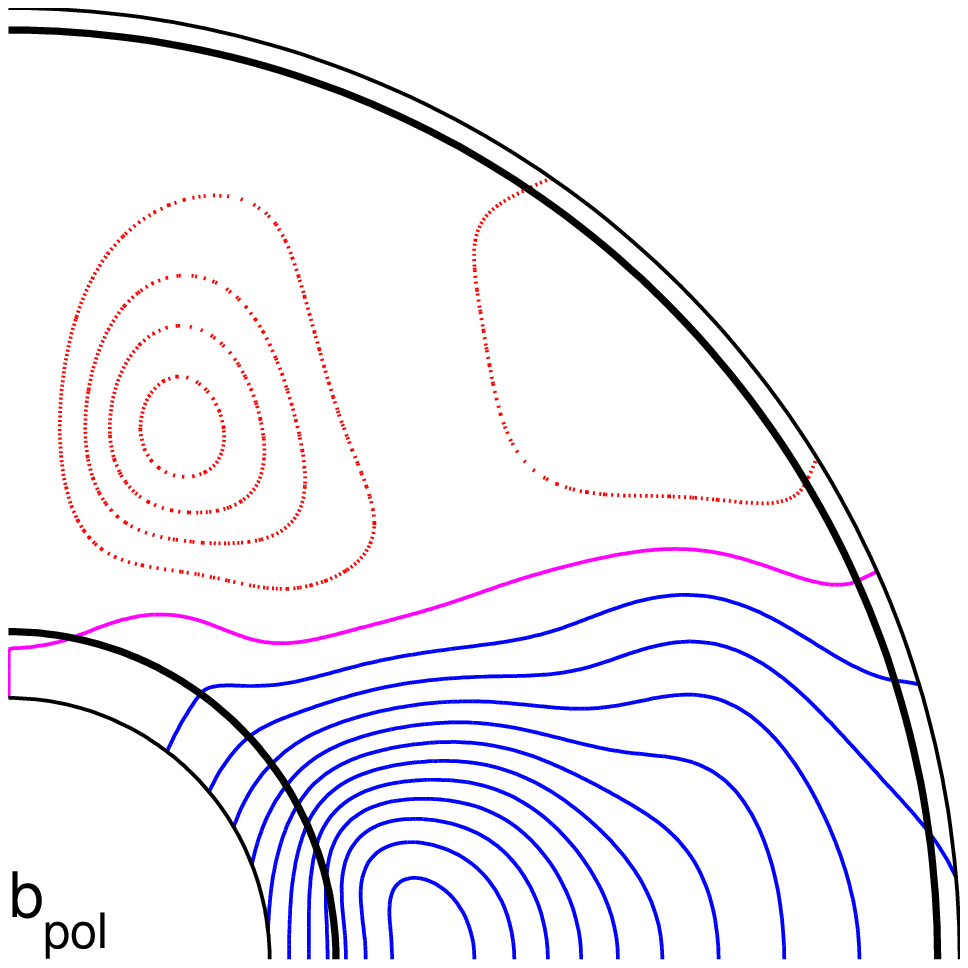} \hspace{-1.2cm} \includegraphics[width=6.5cm]{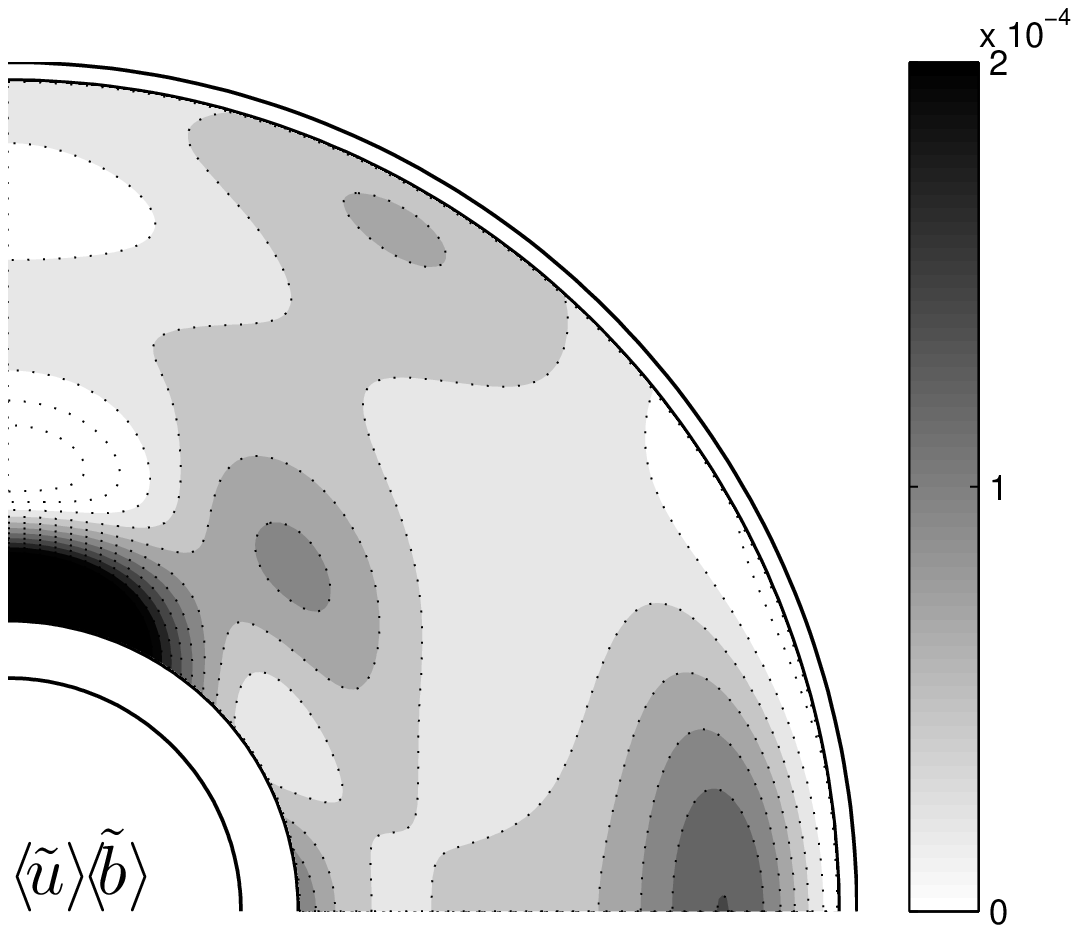}}
	\caption{{\it (left and center)} maps of the contributions of the additional induction terms from equations \ref{eq:alpha} to the azimuthal magnetic field $b_{\varphi}$ and to the meridional field $\boldsymbol{b}_p$.
	The values of the magnetic stream function for $\boldsymbol{b}_p$ range from $-1.2 \; 10^{-6}$ to $0.5 \; 10^{-6}$.
	{\it (right)} tentative map of the product of average fluctuations in velocity and magnetic field $\langle\tilde{u}\rangle\langle\tilde{b}\rangle$.
}
	\label{fig:add}
\end{figure}

Compared with equation \ref{eq:az:induction} used for the inversion, the equation for $b_{\varphi}$ contains three additional terms.
The first one corresponds to the $\omega$-effect applied to the induced meridional magnetic field.
The second one is the advection of the azimuthal field by the meridional flow.
The third one describes the mean-field axisymmetric induction caused by the correlated non-axisymmetric fluctuations of velocity $\tilde{\boldsymbol{u}}$ and magnetic field $\tilde{\boldsymbol{b}}$.
We can compute the first two coupling terms by combining the results of our azimuthal and meridional inversions.
The products are computed in the physical space, and projected onto the $P_n^1$ Legendre polynomials (see \ref{parity and truncation}
for the parity and bounds of the projection).
Solving the induction equation for these two coupling terms, I get the map of the azimuthal magnetic field $b_\varphi$ they produce (figure \ref{fig:add}).
Its amplitude amounts to about $10^{-4} \, \Rm$, yielding a contribution of less than $1\%$ to $b_\varphi$ for $\Rm=9.4$ (compare with figure \ref{fig:az:maps}).

Only one coupling induction term appears in the equation for the meridional magnetic potential $a$.
It corresponds to the advection of the induced meridional field by the meridional flow.
I compute it and solve the induction equation to get the field lines of its contribution to the meridional magnetic field, shown in figure \ref{fig:add}.
Its amplitude is about $10^{-6} \, \Rm$, which modifies the inverted meridional magnetic field of figure \ref{fig:mer:maps} by about $1\%$ for $\Rm=9.4$.

I now turn to the last term of both equations, which is the most interesting one because it describes the contribution of non-axisymmetric fluctuations to the mean axisymmetric magnetic field.
This effect is supposed to play a major role in the generation of large-scale magnetic fields by dynamo action.
There exists a vast literature on the axisymmetric mean-field dynamo equations, which have proven very useful in deciphering the magnetic behaviour of the Sun and stars.
The excellent review by \citet{charbonneau05} recalls that the electromotive force $\mathcal{E} = \langle\tilde{\boldsymbol{u}} \times \tilde{\boldsymbol{b}}\rangle$ term is often expressed as a truncated series expansion of the large-scale magnetic field: $\mathcal{E} = \boldsymbol{\alpha} : \langle\boldsymbol{B}\rangle + \boldsymbol{\beta} : \bnabla \times \langle\boldsymbol{B}\rangle$, where $\boldsymbol{\alpha}$ and $\boldsymbol{\beta}$ are two tensors, which depend upon the turbulent characteristics of the flow.
The $\alpha$-effect is crucial for transferring energy from the azimuthal magnetic field to the meridional field, while the $\beta$-effect can be seen as enhancing the magnetic diffusivity.
Although some theories can provide expressions for these two effects in idealized situations such as homogeneous isotropic turbulence \citep{steenbeck66} or magnetohydrodynamic turbulence \citep{pouquet76}, or phase fluctuations of simple flows \citep{petrelis06}, it remains a challenge to estimate their contribution in actual experiments and in celestial objects.
Recent experiments have shed some light on this issue:

The torus experiment in Perm consists in spinning at rotation rates up to $45$ Hz a torus filled with liquid sodium. The torus is then stopped abruptly, generating a highly turbulent decaying flow.
Using this device, \citet{frick10} have measured the electromagnetic response of the torus channel to an alternating magnetic field, thereby measuring the effective global magnetic diffusivity.
They have observed an enhanced diffusivity during the turbulent decay, which they attribute to a $\beta$-effect.
The magnetic diffusivity was found to be up to $30\%$ larger than the molecular diffusivity, for the highest magnetic Reynolds number $\Rm \simeq 30$.
Local velocities were measured in the same experiments, enabling \citet{noskov12} to compare turbulent magnetic diffusivity with turbulent viscosity, and to propose a scaling for turbulent diffusivity as: $\beta \sim \Rm_{rms}^k$, where $\Rm_{rms} = u_{rms} r_o/\eta$ is a magnetic Reynolds number based on the amplitude of the velocity fluctuations $u_{rms}$, and $r_o$ is the channel radius.
The exponent $k$ is about $2$ for  $\Rm_{rms} < 1$, decreasing to about $1.3$ above that.
A maximum value of $\Rm_{rms} \simeq 1.5$ was achieved in this experiment.

The Madison liquid sodium experiment has two counter-rotating impellers facing each other in a sphere filled with 520 liters of liquid sodium.
With a maximum rotation rate of $30$ Hz, magnetic Reynolds numbers up to $160$ are achieved.
\citet{spence06} have reported a turbulent-induced large-scale magnetic field in this set-up.
Fluctuations induce some $\alpha$-effect, which decreases the induction produced by the mean large-scale flow, thereby preventing dynamo onset.
More recently, \citet{rahbarnia12} have installed a probe that measures the velocity and magnetic field vectors simultaneously at a single position.
This gives access to components aligned with the mean magnetic field (akin to an $\alpha$-effect) and to those aligned with the mean electric current (akin to a $\beta$-effect).
\citet{rahbarnia12} find that the latter dominates in the current set-up that includes an equatorial baffle, which reduces the amplitude of the largest-scale turbulent eddies.
For their highest magnetic Reynolds number $\Rm^{tip}=160$, they measure an excess magnetic diffusivity of about $\beta \simeq 30\%$.
As \citet{noskov12} they introduce a fluctuation-based magnetic Reynolds number: $\Rm_{rms}^* = u_{rms} \ell/\eta$, which differs because the typical dimension $\ell$ is the correlation length of the fluctuations, which they find to be at least one order of magnitude smaller than the integral scale.
The maximum value they reach is $\Rm_{rms}^* \simeq 1.2$, corresponding to $\Rm_{rms} \simeq 12$.
The relation $\beta \simeq \Rm_{rms}^*$ looks compatible with their data.

Getting back to $DTS$ and to equations \ref{eq:alpha}, I note that, unlike the coupling terms, the induction due to non-axisymmetric fluctuations cannot be computed from my inversion results.
In fact, we have no access to the $\langle\tilde{\boldsymbol{u}} \times \tilde{\boldsymbol{b}}\rangle$ term.
However, we can estimate a reasonable upper bound from the product $\langle\tilde{u}\rangle\langle\tilde{b}\rangle$.
In the example shown by \citet{rahbarnia12}, the average electromotive force amounts to about $60\%$ of this product.
Figure \ref{fig:add} displays a tentative map of $\langle\tilde{u}\rangle\langle\tilde{b}\rangle$ in the $DTS$ experiment.
Unlike the other maps in this article, it does not result from a rigorous inversion.
I have assumed a decomposition on $P_n^0$ (with even $n$) of the average velocity and magnetic fluctuations.
I thus constructed a map of the velocity fluctuations from the inversion of the root-mean square fluctuations measured along the same Doppler profiles as for the mean-flow inversions.
Similarly, a map of the magnetic fluctuations was built by inversion of the $rms$ fluctuations of the magnetic field in the sleeve at the three different latitudes.
The map shown in figure \ref{fig:add} is the product of these two maps.

The lack of directional and time-correlation information precludes using this map to solve for the induced magnetic field directly, but we can compare the amplitude of the estimate of this induction term with that of the other additional terms discussed above.
The amplitude of $\langle\tilde{u}\rangle\langle\tilde{b}\rangle$ reaches about $10^{-4} \, \Rm$, while the amplitude of the additional coupling terms is about $2 \,10^{-2} \, \Rm$ in the $b_\varphi$ equation and $5 \, 10^{-4} \, \Rm$ in the equation for $a$.
Therefore, non-axisymmetric fluctuations can contribute at best to a few percents of the total induction at our highest magnetic Reynolds number ($\Rm \simeq 100$), and should always remain smaller than the terms that couple the meridional and azimuthal flows and magnetic fields.

It may look surprising that fluctuations have so little effect in our case, while we reach magnetic Reynolds number larger than \citet{noskov12}.
We know that velocity fluctuations are hampered by the presence of the strong imposed magnetic field \citep{figueroa12}.
Nevertheless, using our estimate of $\tilde{e}_K$ in table \ref{tab:energies}, we get $\Rm_{rms} \simeq 4.5 \, 10^{-2} \, \Rm$, yielding a value $\Rm_{rms} \simeq 4$ at our highest rotation rate.
This is at least twice larger than the maximum value of \citet{noskov12}, but comparable to the lowest value reported by \citet{rahbarnia12} ($\Rm_{rms} \simeq 3.4$ for $\Rm^{tip}=60$), for which the $\beta$-effect is only $2\%$.

However, one should keep in mind that I have focused on the inversion of data obtained at low magnetic Reynolds number ($\Rm=9.4$), and assumed that the results can be extrapolated to our maximum $\Rm$ because \citet{brito11} observe that velocity and induced magnetic field are proportional to $\Rm$ at first order.
But they also observe some important differences when $\Rm$ increases and causes the relative strength of the imposed dipolar field to weaken: the zone of super-rotation shrinks and the intensity of the induced field increases more rapidly with $\Rm$.
We will have to investigate the role of these modifications.
It remains that I was not able to perfectly fit all data at $\Rm=9.4$.
In particular, my model does not predict high enough values for $b_\varphi$, while it overestimates the electric potential differences measured at the surface.
It could due to the fact that the experiment deviates somewhat from the assumptions used: the imposed magnetic field is not perfectly dipolar, the inner sphere is held by shafts, the equatorial symmetry could be slightly violated, etc. 

\citet{nataf08} note that fluctuations get even smaller when both the Coriolis and the Lorentz forces are dominant, as evidenced from measurements in the $DTS$ experiment when the outer sphere spins.
It would be interesting to investigate the role of fluctuations when these constraints are still present but at a much larger magnetic Reynolds number, such as can be achieved in the $3$m-diameter {\it BigSister} experiment of Dan Lathrop and his group at the University of Maryland.

\vspace{12pt}
\noindent
{\bf acknowledgments}
\vspace{6pt}

I thank D. Jault for his guidance throughout this study.
This manuscript was greatly improved by appreciated comments from an anonymous referee and from the Editor.
I thank N. Plihon for useful comments.
I am very grateful to all past and present members of the geodynamo team for their active participation in the $DTS$ adventure.
I gratefully acknowledge the support of CNRS and Universit\'e de Grenoble through the collaborative program ``Turbulence, Magnetohydrodynamics and Dynamo''.

\newpage
\appendix

\section{Adimensionalization}
\label{adimensionalization}

\begin{table}[h] 
 	\begin{center}
		\begin{tabular}{cc}
	quantity & scale \\
	\hline
	time & $\Omega^{-1}$ \\
	length & $r_o^*$ \\
	velocity & $r_o^* \Omega = \frac{\eta}{r_o^*} \Rm$ \\
	imposed magnetic field & $B_o$ \\ 
	induced magnetic field & $\Rm \, B_o$ \\ 
	electric field & $\Rm \,  \eta B_o / r_o^*$ \\
	electric potential & $\Rm \, \eta B_o$ \\
	electric current density & $\Rm \, \eta \sigma B_o / r_o^*$ \\
	magnetic torque & $\Rm \, (r_o^*)^3 B_o^2 / \mu_0$ \\
	energy & $\rho \Omega^2 (r_o^*)^5 = \rho \eta r_o^* \Rm^2$ \\
		\end{tabular}
		\caption{\label{tab:adim}Adimensionalization used in this article. $\Omega$ is the imposed angular velocity of the inner sphere, and $r_o^*$ is the dimensional inner radius of the outer shell. $B_o$ is the intensity of the imposed dipolar magnetic field at the equator for $r=r_o$. $\Rm$ is the magnetic Reynolds number defined as $\Rm = (r_o^*)^2 \, \Omega /\eta$.
The magnetic diffusivity, the electric conductivity, and the density of the liquid are noted $\eta$, $\sigma$ and $\rho$, respectively.}
	\end{center}
\end{table}


\section{Parity and truncation}
\label{parity and truncation}


\begin{table}[h] 
 	\begin{center}
		\begin{tabular}{cccc}
			(r,$\theta$) field & degree parity & minimum degree & maximum degree \\
			\hline
			$U_\varphi$ & odd & 1 & $l_{max}$ \\
			$b_\varphi$ & even & 2 & $l_{max} + 1$ \\
			$E_\theta$ & even & 2 & $l_{max} + 1$ \\
			$\mathcal{F_\varphi}$ & odd & 1 & $l_{max} + 2$ \\
			\hline
			$u$ & even & 2 & $n_{max}$ \\
			$a$ & odd & 1 & $n_{max} + 1$ \\
			\hline
			coupling terms for $b_\varphi$ & even & 2 & $l_{max}+n_{max}-1$ \\
			coupling terms for $a$ & odd & 1 & $2*n_{max}-1$ \\
	\end{tabular}
		\caption{\label{tab:truncation}Parity and truncation degree of the various ($r$,$\theta$) fields analyzed in this article. Azimuthal inversion: $U_\varphi$, $b_\varphi$, $E_\theta$ and $\mathcal{F_\varphi}$; meridional inversion: $u$ and $a$; coupling terms in equation \ref{eq:alpha} for $b_\varphi$ and $a$.}
	\end{center}
\end{table}

\section{Dealing with discontinuities in velocity and conductivity}
\label{solid shells}

In this appendix, I give the analytic expressions of the magnetic and electric fields in the solid spherical shells that contain the liquid sodium.
I take into account the different electric conductivities and the contribution of a discontinuity of the tangential velocity at the interface between the solid and the fluid.
I take advantage of the fact that we only consider the steady-state solution. 

\subsection{magnetic field}
\label{magnetic field}

In the solid inner and outer shells, equation (\ref{eq:az:induction}) reduces to $\Delta_1b_{\varphi}=0$, which implies that the $b_n(r)$ functions of the decomposition (\ref{eq:az:Legendre_b}) satisfy:

\begin{equation}
\frac{1}{r} \frac{d^2 (r b_n)}{d r^2} - \frac{n(n+1)}{r^2} b_n = 0,
\label{eq:diffusion_in_solid}
\end{equation}
whose solutions are written as:

\begin{equation}
^sb_n(r) = c_n^+ r^n + c_n^- r^{-(n+1)},
\label{eq:solution_in_solid}
\end{equation}
where the $^s$ left superscript refers to the solid shell.
Since electric currents cannot circulate outside the shells, $b_{\varphi} = 0$ at $r=\hat{r}_i$ and $r=\hat{r}_o$.
Adding the continuity of $b_{\varphi}$ across the boundaries, we obtain the following relations between the $^+$ and $^-$ coefficients of equation (\ref{eq:solution_in_solid}):

\begin{equation}
\left \{ \begin{array}{rl}
c_n^{+} \hat{r}_i^n + c_n^{-} \hat{r}_i^{-(n+1)} & = 0 \\
c_n^{+} r_i^n + c_n^{-} r_i^{-(n+1)} & = b_n(r_i).
\end{array} \right.
\label{eq:solid_bc}
\end{equation}
Solving this linear system, we obtain:
\begin{eqnarray}
c_n^+ &=& - (\hat{r}_i^{-(n+1)}/D_n) \; b_n(r_i) \\
c_n^- &=& (\hat{r}_i^n / D_n) \; b_n(r_i),
\label{eq:c}
\end{eqnarray}
where the determinant $D_n$ is:
\begin{equation}
D_n = \hat{r}_i^n r_i^{-(n+1)} - \hat{r}_i^{-(n+1)} r_i^n.
\label{eq:determinant}
\end{equation}

The continuity of the $\theta$ component of the electric field yields the final relation.
On the two sides of the solid/fluid interface, we have:

\begin{equation}
\left \{ \begin{array}{rl}
^sE_{\theta} &= - \frac{^l\sigma}{^s\sigma} \left. \frac{\partial (r \; ^sb_{\varphi})}{r \partial r} \right|_s - \; ^sU_{\varphi} B_r \\
^lE_{\theta} &= - \left. \frac{\partial (r b_{\varphi})}{r \partial r} \right|_l - \; U_{\varphi} B_r ,
\end{array} \right.
\label{eq:E_continuity}
\end{equation}
where the $^l$ left superscript refers to the liquid.
The continuity of $E_{\theta}$ thus yields the following relation between the jumps in velocity and conductivity, and that in the derivative of the magnetic field at $r_i$ (with our adimensionalization given in \ref{adimensionalization}):

\begin{equation}
\left. \frac{\partial (r b_{\varphi})}{r \partial r} \right|_{r_i} = \frac{^{Na}\sigma}{^{Cu}\sigma} \left. \frac{\partial (r \; ^{Cu}b_{\varphi})}{r \partial r} \right|_{r_i} - \left[ U_{\varphi}(r_i,\theta) - \; ^{Cu}U_{\varphi}(r_i,\theta) \right] B_r(r_i,\theta),
\label{eq:b_derivative_bc}
\end{equation}
where $^{Na}\sigma$ and $^{Cu}\sigma$ are the electrical conductivities of liquid sodium and of the copper inner shell, respectively.
The azimuthal velocity of the inner sphere is $^{Cu}U_{\varphi}(r_i,\theta) = r_i \, \sin\theta = -r_i \, P_1^1(\cos\theta)$.
We now evaluate the derivative term of equations \ref{eq:b_derivative_bc} in the solid part, using expression \ref{eq:solution_in_solid}:
\begin{equation}
\left. \frac{\partial (r \; ^{Cu}b_{\varphi})}{r \partial r}  \right|_{r_i} = \sum_{even \; n} \mathcal{C}_n \; b_n(r_i) \; P_n^1(\cos \theta),
\label{eq:solid_derivative}
\end{equation}
with:
\begin{equation}
\mathcal{C}_n = -\frac{(n+1) r_i^{n-1} \hat{r}_i^{-(n+1)} + n \, r_i^{-(n+2)} \hat{r}_i^n}{D_n}.
\label{eq:C}
\end{equation}
Injecting equations \ref{eq:solid_derivative} in \ref{eq:b_derivative_bc} and expressing the radial derivative of $rb_{\varphi}$ in the fluid, we get:
%
%
\begin{equation}
\sum_{even \; n} \left [ \left ( \frac{1}{r_i} - \frac{^{Na}\sigma}{^{Cu}\sigma} \mathcal{C}_n \right ) b_n(r_i) + \left. \frac{d \, b_n}{dr} \right|_{r_i} \right ] P_n^1(\cos \theta)
= - \left[ U_{\varphi}(r_i,\theta) - r_i \sin\theta \right] \; B_r(r_i,\theta).
\label{eq:derivative_1}
\end{equation}
Making use of the Legendre projection (\ref{eq:E_theta}) for the $u \times B$ term, we finally obtain:
\begin{equation}
\begin{split}
\sum_{even \, n} \left [ \left ( \frac{1}{r_i} - \frac{^{Na}\sigma}{^{Cu}\sigma} \mathcal{C}_n \right ) b_n(r_i) + \left. \frac{d \, b_n}{dr} \right|_{r_i} \right ] P_n^1 \\
= -\frac{2}{r_i^3}\sum_{odd \; l} \left( U_l(r_i) + r_i \, \delta_{1 l} \right) \left[ \frac{l+1}{2l+1}P_{l-1}^1+\frac{l}{2l+1}P_{l+1}^1\right],
\end{split}
\label{eq:derivative_2}
\end{equation}
%
%
where $\delta_{1 l}$ is the Kronecker symbol.
By identification of the $P_l^1$ of identical degree, we deduce a relationship between the magnetic field and its derivative at the interfaces for each even degree $2 \le n \le l_{max}+1$:

\begin{equation}
\left. \frac{d \, b_n}{dr} \right|_{r_i} = - \mathcal{P}_n \, b_n(r_i) - \mathcal{Q}_n,
\label{eq:derivative_3}
\end{equation}
with:
\begin{eqnarray}
\mathcal{P}_n &=& \frac{1}{r_i} - \frac{^{Na}\sigma}{^{Cu}\sigma} \mathcal{C}_n \\
\mathcal{Q}_n &=& \frac{2}{r_i^3} \left( \frac{n+2}{2 n + 3} U_{n+1}(r_i) + \frac{n-1}{2 n - 1} (U_{n-1}(r_i) + r_i \, \delta_{n2})  \right).
\label{eq:derivative_4}
\end{eqnarray}

This condition is similar (albeit more complex) to the one encountered for the magnetic potential induced by the meridional circulation.
The boundary condition at the interface between the fluid and the outer shell is exactly the same, replacing $r_i$ by $r_o$ and $\hat{r}_i$ by $\hat{r}_o$, $^{Cu}\sigma$ by $^{SS}\sigma$ (the electric conductivity of the stainless steel outer shell), and omitting the $\delta_{n2}$ term in $\mathcal{Q}_n$ since the outer sphere is at rest.

%

\subsection{electric field}
\label{electric field}

Using the expression \ref{eq:E_continuity} of the $\theta$ component of the electric field in the solid shell, and injecting the decomposition \ref{eq:solution_in_solid} of the $b_n$ functions, we obtain an analytical expression for the $e_n$ functions in the inner copper shell:
\begin{equation}
\begin{split}
^{Cu}e_n(r) = \frac{^{Na}\sigma}{^{Cu}\sigma} \; \left[ (n+1) \hat{r}_i^{-(n+1)} r^{n-1} + n \, \hat{r}_i^n \, r^{-(n+2)}\right] \frac{b_n(r_i)}{D_n} +\frac{2}{3 r^2}\delta_{n2}.
\end{split}
\label{eq:e_p_s}
\end{equation}
A similar relation is easily derived for the outer stainless steel shell.

\newpage

\bibliographystyle{elsarticle-harv}
\bibliography{biblio}







\end{document}